\documentclass{ptephy}

\usepackage{boites}
\usepackage{amsmath,amssymb}
\usepackage{graphicx}
\usepackage{float}
\usepackage{tabularx}
\usepackage{xcolor}

\def\beq{\begin{equation}}
\def\eeq{\end{equation}}
\def\be{\begin{equation}}
\def\ee{\end{equation}}
\def\bea{\begin{eqnarray}}
\def\eea{\end{eqnarray}}
\def\ba{\begin{array}}
\def\ea{\end{array}}
\def\<{\left\langle}
\def\>{\right\rangle}
\def\({\left(}
\def\){\right)}
\def\e{{\rm e}}
\def\tr{{\rm tr}}
\def\nn{\nonumber}
\def\openone{\mathbb{I}}
\newcommand\bs[1]{\boldsymbol{\mathit{#1}}}

\begin{document}

\title{
Individual eigenvalue distributions of crossover chiral random matrices
and low-energy constants of SU(2)$\times$U(1) lattice gauge theory}

\author{
\name{\fname{Takuya} \surname{Yamamoto}}{}
and 
\name{\fname{Shinsuke M.} \surname{Nishigaki}}{\ast}
} 

\address{\affil{~}{Graduate School of Science and Engineering, Shimane University, Matsue 690-8504, Japan}
\email{mochizuki@riko.shimane-u.ac.jp}}

\begin{abstract}%
We compute individual distributions of low-lying eigenvalues
of a chiral random matrix ensemble interpolating
symplectic and unitary symmetry classes by
the Nystr\"{o}m-type method of evaluating the Fredholm Pfaffian and 
resolvents of the quaternion kernel.
The one-parameter family of these distributions is shown to fit excellently
the Dirac spectra of
SU(2) lattice gauge theory with a constant U(1) background or dynamically fluctuating U(1) gauge field,
which weakly breaks the pseudoreality of the unperturbed SU(2) Dirac operator.
The observed linear dependence of the crossover parameter with the strength of the U(1) perturbations
leads to precise determination of the pseudo-scalar decay constant, as well as the chiral condensate
in the effective chiral Lagrangian of the AI class.
\end{abstract}

\subjectindex{B01,B60,B64,B83,B86}

\maketitle

\section{Introduction}
The grounds for universality, i.e., insensitivity to details of the system of concern, 
of local correlation of energy levels of stochastic and quantum-chaotic Hamiltonians
have been well uncovered by now,
in terms of ten-fold classification of symmetric superspaces
on which spectral nonlinear $\sigma$ models describing spontaneous symmetry breaking reside \cite{Zir96}, 
and of semiclassical equivalence between periodic orbits and the aforementioned $\sigma$ models \cite{MHABH09}. 
On the other hand, the presence of explicit symmetry breaking perturbations is known to induce
a crossover between different universality classes \cite{Dys62},
in such a way that is also insensitive to the systems' details.
An example of this universality crossover is the Gaussian orthogonal ensemble (GOE)--Gaussian unitary ensemble 
(GUE) transition in a disordered or chaotic system under a magnetic field \cite{DM91,SNMB09}.
In the realm of lattice gauge theory, where Dirac operators play the r\^{o}le of stochastic Hamiltonians \cite{SV93,Ver94},
the crossover between the chiral Gaussian unitary ensemble (chGUE) itself and the chGUE--GUE crossover,
associated with an imaginary isospin chemical potential \cite{DHSS05} and
a finite lattice-spacing effect in the Wilson Dirac operator \cite{DSV10}, respectively,
have been utilized to
determine the pion decay constant and the Wilsonian chiral perturbation constants
from relatively small lattices.
The aim of this work is to apply this spectral approach to the determination of low-energy constants 
in another setting, namely SU(2) gauge theory under U(1) perturbations, either in the form of 
a constant imaginary chemical potential \cite{Nis12-1,Nis12-2} or a dynamically fluctuating one. 
In contrast to these preceding works which used $n$-level correlation functions or the
smallest eigenvalue distribution,
our strategy in this paper is to employ {\em multiple} spectral observables
which allows for precise fitting of lattice data,
namely, individual distributions of the $k$th smallest Dirac eigenvalues \cite{DN01} inclusively.
The practical advantages of our method will be proved in the precision of the low-energy constants determined,
as preliminarily reported in Ref.~\cite{NY14}.

This paper is composed mainly of two parts: 
Sect.~2 is devoted to analytic treatments in random matrix theory, and 
Sect.~3 to its application to Dirac eigenvalue distributions measured in lattice simulations.
In Sect.~2 we start by reviewing the established results of the chiral Gaussian symplectic ensemble 
(chGSE)--chGUE crossover,
namely, the derivation of the quaternion kernel $K$ \cite{FNH99,NF99,Nag05}.
Then we apply the Nystr\"{o}m-type method \cite{Bor10-1,Bor10-2} to that kernel and compute
individual eigenvalue distributions in the form of the  Fredholm Pfaffian and resolvents of $\hat{K}$ \cite{Nis15}.
The relationship between the chiral Lagrangian in the $\varepsilon$ regime
and
the nonlinear $\sigma$ model from random matrices interpolating chGSE--chGUE 
leads to identification between parameters in each theory.
In Sect.~3 we introduce our models of lattice gauge theory and
explain our strategy of fitting the Dirac spectra using individual eigenvalue distributions
of chGSE and of the chGSE--chGUE crossover.
Optimally fitting parameters (mean level spacings $\varDelta$ and crossover parameters $\rho$)
will be exhibited in Tables 1--6, leading to very precise determination of
the low-energy constants (chiral condensate $\Sigma$ and pseudo-scalar decay constant $F$)
as presented in Table 7.
Our conclusions, including a possible direction of study, will be summarized in Sect.~4.

\section{chGSE--chGUE crossover}

\subsection{Crossover random matrix ensemble}
Let $N$ and $N'$ be even positive integers.
Let $\bs{A}$ and $\bs{B}$ be $(N/2)\times (N'/2)$ quaternion matrices,
which can be represented as ordinary $N\times N'$ matrices $A$ and $B$ as
\footnote{
We denote quaternions
in bold symbols and their $2\times 2$ complex matrix representatives in the corresponding italic symbols.}
\be
A=\sum_{\mu=0}^3 \(A^{(\mu)}_{jk}\) \otimes \sigma_\mu,\ \  
B=\sum_{\mu=0}^3 \(B^{(\mu)}_{jk}\) \otimes \sigma_\mu\ \ 
(j=1,\ldots,N/2,\ k=1,\ldots,N'/2).
\ee
Here the four units of
the quaternion field $\mathbb{H}$
are represented by the $2\times 2$ unit matrix and Pauli matrices,
$\{\sigma_\mu \}=(\openone, -i\sigma_1, -i\sigma_2,- i\sigma_3)$.
Let these matrix elements belong to
\be
A^{(\mu)}_{jk}\in \mathbb{R},\ B^{(\mu)}_{jk}\in \mathbb{C},
\ee
i.e.,~$\bs{A}$ is quaternion-real and $\bs{B}$ is not (i.e., $B$ is a generic $N\times N'$ complex matrix).
We consider
$A^{(\mu)}_{jk}$, ${\rm Re}\, B^{(\mu)}_{jk}$, and ${\rm Im}\, B^{(\mu)}_{jk}$
to be independent random variables,  distributed according to
the Gaussian distributions
$e^{-\frac12\tr\, A A^\dagger}$ and $e^{-\tr\, B B^\dagger}$, respectively.
We define an ensemble of $(N+N') \times (N+N') $ Hermitian matrices $H$ of the form
\be
H=
\left[
\ba{cc}
0_{N\times N} &C\\
C^{\dagger} &0_{N'\times N'}
\ea
\right],
\ \ C = e^{-\tau}A+\sqrt{1-e^{-2\tau}} B,
\label{H}
\ee
where $\tau$ is a real parameter, initially introduced by Dyson as a fictitious time for the Brownian motion
of eigenvalues \cite{Dys62}. 
This ensemble is called a ``chiral'' random matrix ensemble because it enjoys
the chiral symmetry $\left\{H,\gamma_5\right\}=0$ with $\gamma_5={\rm diag}(\openone_{N},-\openone_{N'})$.
This anticommutation relation
implies that
the eigenvalues of $H$ consist of 
$\min (N',N)$ pairs of generically nonzero eigenvalues of equal magnitude and opposite signs
(i.e., $(\pm1) \times$ singular values of $C$),
and $\nu=|N'-N|$ zero eigenvalues.
The presence of $B$ violates the quaternion-reality of $\bs{C}$
and the self-duality of $H$ 
(i.e., \ $H_{ij}^{(\mu)}\otimes\sigma_\mu=H_{ji}^{(\mu)}\otimes\sigma_\mu^\dagger,\ i,j=1,\ldots,N+N'$),
and lifts the Kramers degeneracy
of all nonzero eigenvalues of $H$ (i.e., \ singular values of $A$ alone).
Accordingly, this random matrix ensemble interpolates between two limiting cases,
the chiral Gaussian Symplectic Ensemble (chGSE) at $\tau=0$ and 
the chiral Gaussian Unitary Ensemble (chGUE) at $\tau\to\infty$,
depending on a single parameter $\tau$ (at a fixed, finite $N$ and $N'$).

\subsection{Joint eigenvalue distribution}
In order to make the paper self-contained,
below we sketch the derivation of the probability distribution of singular values of $C$,
and refer the reader to Refs.~\cite{FNH99,NF99,Nag05} for rigorous proofs of the relevant formulas. 
We start from the unnormalized probability measure of the matrix elements of $C$,
\bea
&&dC\, 
\int dA\, e^{-\frac12 \tr\, A A^\dagger}
\int dB\, e^{- \tr\, B B^\dagger}
\delta( e^{-\tau}A+\sqrt{1- e^{-2\tau}} B-C)
\nonumber\\
&\propto&
dC\,
\int dA\,\exp\(
-\frac12 \tr\, A A^\dagger
-\frac{1}{1- e^{-2\tau}}\tr\, (C- e^{-\tau}A) (C- e^{-\tau}A)^\dagger
\),
\label{dC}
\eea
where $dA=\prod_{j,k,\mu}dA_{jk}^{(\mu)}$, $dB=\prod_{j,k,\mu}d^2 B_{jk}^{(\mu)}$, and 
$dC=\prod_{j,k,\mu}d^2 C_{jk}^{(\mu)}$.
Without loss of generality we assume $N\leq N'$.
We employ the singular value decomposition
$A=S M S'{}^\dagger$ and $C=U \Lambda U'{}^\dagger$
with $S\in {\rm USp}(N), S'\in {\rm USp}(N'), U\in {\rm U}(N), U'\in {\rm U}(N')$ 
and parametrize the singular values as
$M=\Bigl[{\rm diag}\(\mu_1,\ldots,\mu_{N}\)\ 0_{N\times\nu}\Bigr],\ 
\Lambda=\Bigl[{\rm diag}\(\lambda_1,\ldots,\lambda_{N}\)\ 0_{N\times\nu}\Bigr]$.
Kramers-degenerate pairs of singular values of $A$ are ordered such that 
$\mu_{i+N/2}=\mu_i\ (i=1,\ldots,N/2)$.
The measures $dA$ and $dB$ on quaternion-real matrices and complex matrices
take the following respective forms (in what follows we suppress all constant factors in the measures): 
\be
dA=d(S, S') \(\prod_{i=1}^{N/2} d\mu_i\,\mu_i^{2\nu+3}\)\triangle_{N/2}(\mu^2)^4,\ \ 
dC=d(U, U') \(\prod_{i=1}^{N} d\lambda_i\,\lambda_i^{2\nu+1}\)\triangle_{N}(\lambda^2)^2.
\label{measures}
\ee
Here $d(S, S')$ and $d(U, U')$ denote the invariant measures on the respective angular degrees of freedom,
and $\triangle$ denote the Vandermonde determinants
$\triangle_{N/2}(\mu^2):=\prod_{i>j}^{N/2} (\mu_i^2-\mu_j^2)$ and 
$\triangle_{N}(\lambda^2):=\prod_{i>j}^{N} (\lambda_i^2-\lambda_j^2)$.
The probability measure of the singular values $\{\lambda_i\}$ of $C$ follows
from Eqs. (\ref{dC}) and (\ref{measures}) by integrating out the unitary matrices $(U, U')$.
The integrations over symplectic matrices $(S, S')$ decouple after redefining the unitary
matrices by  $S^\dagger U\to U$ and $S'{}^\dagger U'\to U'$, leading to the expression
\bea
\text{Eq.} (\ref{dC})&=&\prod_{i=1}^{N} d\lambda_i\,\lambda_i^{2\nu+1}\,\exp\(
-\frac{\lambda_i^2}{1- e^{-2\tau}}
\)
\triangle_{N}(\lambda^2)^2
\int_0^\infty 
\prod_{j=1}^{N/2} d\mu_j\,\mu_j^{2\nu+3}\,\exp\(
-\frac{\mu_j^2}{\tanh \tau}
\)
\nonumber\\
&&\times
\triangle_{N/2}(\mu^2)^4
\int_{{\rm U}(N)}  \!\!\!\!\!\!\! dU
\int_{{\rm U}(N')} \!\!\!\!\!\!\!\!\! dU' \ 
\exp\(
\frac{1}{\sinh \tau}{\rm Re}\,\tr\,U\Lambda U'{}^\dagger M^t
\).
\label{dlambda} 
\eea
We employ the Berezin--Karpelevich formula \cite{BK68,GW96,JSV96} for the integration over $(U, U')$
and take the pairwise confluent limit $\mu_{i+N/2}\to \mu_i$ for all $i=1,\ldots, N/2$:
\bea
\int_{{\rm U}(N)}  \!\!\!\!\!\!\!\! dU  
\int_{{\rm U}(N')} \!\!\!\!\!\!\!\!\!\! dU' 
\ \; \, e^{\frac{1}{\sinh\tau}{\rm Re}\,\tr\, U \Lambda U'{}^\dagger M^t}
&\propto&\left.
\frac{\det
\left[ I_{\nu}\(\frac{\lambda_i \mu_j}{\sinh\tau}\) \right]_{i,j=1}^{N}
}{
\triangle_{N}(\lambda^2)\triangle_{N}(\mu^2)
\prod_{i=1}^{N}(\lambda_i \mu_i)^{\nu}}
\right|_{\mu_{i+N/2}\to \mu_i}
\nonumber\\
&\propto&
\frac{\det
\left[ I_{\nu}\(\frac{\lambda_i \mu_j}{\sinh\tau}\) \ 
\frac{\partial}{\partial \mu_j^2}I_{\nu}\(\frac{\lambda_i \mu_j}{\sinh\tau}\)  \right]_{i=1,\ldots,N}^{j=1,\ldots, N/2}
}{\triangle_{N}(\lambda^2)\triangle_{N/2}(\mu^2)^4\prod_{i=1}^{N}\lambda_i^{\nu} \prod_{j=1}^{N/2} 
\mu_j^{2\nu}}.
\label{dlambda1}
\eea
Here $I_\nu$ denotes the Bessel function of the pure imaginary argument, $I_\nu(z)=J_\nu(i z)$.
By substituting Eq. (\ref{dlambda1}) into Eq. (\ref{dlambda}) and performing a change of
variables from the singular values of rectangular matrices $C$ and $A$
to the eigenvalues of Wishart matrices $C C^\dagger$ and $A A^\dagger$,
$x_i=\lambda_i^2$ and $y_i=\mu_i^2$,
the probability measure becomes proportional to
\be
\text{Eq.} (\ref{dlambda})\ \propto\ 
\prod_{i=1}^{N} dx_i\,
\sqrt{w(x_i)}\,
\triangle_{N}(x)
\int_0^\infty 
\prod_{j=1}^{N/2} dy_j\,y_j
\det
\left[
g(x_k, y_\ell) \ \ \frac{\partial g(x_k, y_\ell)}{\partial y_\ell}  \right]^{k=1,\ldots,N}_{\ell=1,\ldots, N/2}.
\label{dlambda2}
\ee
Here we have introduced the Laguerre weight $w(x):=x^{\nu} e^{-x}$, and 
the symmetric function
\be
g(x,y):= \frac{e^{-(2\nu+1)\tau}}{1-e^{-2\tau}}\exp\(-\frac{x+y}{2\tanh \tau}\) I_\nu\( \frac{\sqrt{x y}}{\sinh\tau}\).
\ee
With the multiplicative constant chosen as the above, the function $g(x,y)$
admits an alternative interpretation as a one-particle Green's function for the Brownian motion at time $\tau$,
\be
g(x,y)=\sqrt{w(x) w(y)}
\sum_{k=0}^\infty \frac{L_k^\nu(x)L_k^\nu(y)}{h_k} e^{-\gamma_k \tau},
\ee
with the norm given by $h_k={(k+\nu)!}/{k!}$ and  the ``one-particle energy'' by $\gamma_k=2k+\nu+1$.
Using a lemma by Mehta (A.17 of Ref.~\cite{Meh04}),
the $(N/2)$-fold integral of an $N\times N$ determinant in Eq. (\ref{dlambda2}) can be decomposed into
an $N\times N$ Pfaffian of single integrals: 
\bea
\text{Eq.} (\ref{dlambda2})&\propto&\prod_{i=1}^{N} dx_i\,
\sqrt{w(x_i)}\,
\triangle_{N}(x)\,
{\rm Pf}
\left[
F(x_j, x_k)
\right]_{j,k=1,\ldots,N}, 
\label{dlambda3}\\ 
F(x,x')&:=&
\int_0^\infty dy\,y
\left\{
g(x, y)\frac{\partial g(x',y)}{\partial y}
-\frac{\partial g(x,y)}{\partial y}g(x', y)
\right\}.
\label{F}
\eea

\subsection{Quaternion determinant}
In this subsection we summarize the procedure presented in Ref.~\cite{Meh04}, Sect. 14.
We introduce a set of arbitrary monic polynomials $\{R_k(x)\}_{k=0,1,\ldots}$
and arbitrary positive numbers $\{r_k\}_{k=0,1,\ldots}$, and
define functions $\{\psi_k(x)\}$ by
\be
\psi_{2k}(x)=\frac{\sqrt{w(x)} R_{2k}(x)}{\sqrt{r_k}},\quad
\psi_{2k+1}(x)=\frac{\sqrt{w(x)} R_{2k+1}(x)}{\sqrt{r_k}}
\label{psi}
\ee
and their $F$-convolutions $\{\phi_k(x)\}$ by
\be
\phi_k(x)=-\int_0^\infty dx' \, F(x,x') \psi_k(x').
\label{phi}
\ee
Next we introduce functions $D(x, x')$, $S(x,x')$, and $I(x,x')$, 
which are bilinear combinations of $\{\psi_k(x)\}$ and $\{\phi_k(x)\}$:
\bea
D(x,x')&=&\sum_{k=0}^{N/2-1}\Bigl(\psi_{2k}(x) \psi_{2k+1}(x')-\psi_{2k+1}(x) \psi_{2k}(x')\Bigr),
\label{Dxx}\\
S(x,x')&=&\sum_{k=0}^{N/2-1}\Bigl(\phi_{2k}(x) \psi_{2k+1}(x')-\phi_{2k+1}(x) \psi_{2k}(x')\Bigr),
\label{Sxx}\\
I(x,x')&=&-\sum_{k=0}^{N/2-1}\Bigl(\phi_{2k}(x) \phi_{2k+1}(x')-\phi_{2k+1}(x) \phi_{2k}(x')\Bigr),
\label{Ixx}
\eea
and the corresponding $N\times N$ matrices $D_N$, $S_N$, and $I_N$ by
\be
{D}_N=[D(x_i, x_j)]_{i,j=1,\ldots,N},\ 
{S}_N=[S(x_i, x_j)]_{i,j=1,\ldots,N},\ 
{I}_N=[I(x_i, x_j)]_{i,j=1,\ldots,N}.
\ee
Since a $2N\times 2N$ antisymmetric matrix
$
\left[
\ba{cc}
D_N & S_N{}^t \\
-S_N & -I_N
\ea
\right]
$
is a product of two rectangular matrices of size $2N\times N$ and $N\times 2N$:
\[
\left[
\ba{cc}
D_N & S_N{}^t \\
-S_N & -I_N
\ea
\right]=
\left[
\ba{cc}
\psi_{2k}(x_i) & \psi_{2k+1}(x_i) \\
-\phi_{2k}(x_i) & -\phi_{2k+1}(x_i)
\ea
\right]^{i=1,\ldots, N}_{k=0,\ldots, N/2-1} 
\left[
\ba{cc}
\psi_{2k+1}(x_j) & -\phi_{2k+1}(x_j) \\
-\psi_{2k}(x_j) & \phi_{2k}(x_j)
\ea
\right]_{j=1,\ldots, N}^{k=0,\ldots, N/2-1},
\]
its rank is $N$ at most; and also it is $N$ at least
due to the linear independence of $\{\psi_i(x)\}_{i=0,\ldots,N-1}$.
Accordingly the lower $N$ rows $\left[ -S_N \ -I_N \right]$
are linear combinations of the upper $N$ rows $\left[ D_N \ S_N{}^t \right]$.
Now consider a Pfaffian of another antisymmetric $2N\times 2N$ matrix
$
\left[
\ba{cc}
D_N & S_N{}^t \\
-S_N & -I_N-F_N
\ea
\right]$
with
${F}_N=[F(x_i, x_j)]_{i,j=1,\ldots,N}$.
Since adding
$\left[ S_N \ I_N \right]$ (or minus its transpose) to the lower $N$ rows (the right $N$ columns)
does not change the determinants
due to the aforementioned linear dependence with the upper $N$ rows (the left $N$ columns),
we readily obtain
\be
{\rm Pf}
\left[
\ba{cc}
D_N & S_N{}^t \\
-S_N & -I_N-F_N
\ea
\right]
=
{\rm Pf}
\left[
\ba{cc}
D_N & 0 \\
0 & -F_N
\ea
\right]
=(-1)^{N/2}\,{\rm Pf}\, D_N\cdot{\rm Pf}\, F_N.
\label{PfDPfF}
\ee
On the other hand,
\bea
&&
{\rm Pf}\, D_N=
{\rm Pf}\,\left[
\left[
\psi_{2k}(x_i)\ \psi_{2k+1}(x_i)
\right]^{i=1,\ldots,N}_{k=0,\ldots,N/2-1}
\left[
\ba{c}
\psi_{2k+1}(x_j)\\ -\psi_{2k}(x_j)
\ea
\right]^{k=0,\ldots,N/2-1}_{j=1,\ldots,N}
\right]
\nn\\
&&=
\det
\left[
\psi_{k-1}(x_i)
\right]_{i,k=1,\ldots,N}
=\det
\left[\sqrt{w(x_i)} x_i^{k-1}
\right]_{i,k=1,\ldots,N}
\propto\prod_{i=1}^N \sqrt{w(x_i)}
\cdot
\triangle_N (x).
\label{PfD}
\eea
Using Eqs. (\ref{PfDPfF}) and (\ref{PfD}),
the probability measure (\ref{dlambda3}) now reads
\bea
&& \text{Eq.} (\ref{dlambda3})
\ \propto\ 
\prod_{i=1}^{N} dx_i\,
{\rm Pf}\(
Z K_N\),
\label{jpd6}
\\
&&
K_N=\left[
\ba{cc}
S_N & J_N\\
D_N & S_N{}^t
\ea
\right],\ \ 
J_N=I_N+F_N,\ \ 
Z=
\left[
\ba{cc}
0 & \openone_N \\
-\openone_N & 0
\ea
\right].
\nonumber
\eea
Since its upper diagonal block is the transpose of the lower diagonal block 
and both off-diagonal blocks are antisymmetric,
$K_N$ can be regarded a $2N\times 2N$ complex matrix representative of an $N\times N$
quaternion self-dual matrix, which we call 
$\bs{K}_N=[\bs{K}(x_i,x_j)]_{i,j=1,\ldots,N}$.
Using Dyson's lemma 
${\rm qdet}\,\mathbf{\Phi}={\rm Pf}\,(Z \Phi)$
for the quaternion determinant (qdet) \cite{Dys70}
of a quaternion self-dual matrix $\mathbf{\Phi}$, we finally obtain
\be
\text{Eq.} (\ref{jpd6})=
\prod_{i=1}^{N} dx_i\,{\rm qdet}\bs{K}_N(x_1,\ldots,x_N) .
\label{jpd7}
\ee

\subsection{Skew-orthogonal polynomials}\label{SOP}
We would like to choose $\{R_k(x)\}$ and $\{r_k\}$ (which are so far arbitrary) in such a way that
the $2\times 2$ matrix
$
K(x,y)=\left[
\ba{cc}
S(x,y) & J(x,y)\\
D(x,y) & S(y,x)
\ea
\right]
$
representing the quaternion kernel $\bs{K}(x,y)$
enjoys the quasi-projectivity
\be
\int_0^\infty dy\,
{K}(x,y){K}(y,z)={K}(x,z)
\left[
\ba{cc}
1 & 0\\
0 & 0
\ea
\right]
+
\left[
\ba{cc}
0 & 0\\
0 & 1
\ea
\right]
{K}(x,z),
\label{qproj}
\ee
and is correctly normalized: 
\be
\int_0^\infty dx\,
{K}(x,x)=N.
\label{qnorm}
\ee
These two relationships would yield a crucial property that the restricted quaternion matrix 
$\bs{K}_n=[\bs{K}(x_i,x_j)]_{i,j=1,\ldots,n}$ satisfies the recursion relation
\be
\int_0^\infty dx_n\,{\rm qdet} \,\bs{K}_{n}(x_1,\ldots,x_n)
=(N-n+1)\,{\rm qdet}\,\bs{K}_{n-1}(x_1,\ldots,x_{n-1}),
\ee
from which a $k$-level correlation function is expressed as 
$\mathrm{qdet}\,\bs{K}_{k}=\mathrm{Pf}(Z K_k)$.
It is well established that the properties (\ref{qproj}), (\ref{qnorm}) are fulfilled by requiring
 $\{R_k(x)\}$ to be skew-orthogonal with respect to the skew inner product $\<~,~\>$
and $\{r_k\}$ to be their skew-norms,
\bea
&&\<f, g\>:=\int_0^\infty dx \int_0^\infty dy \, \sqrt{w(x)w(y)}F(x,y)f(x)g(y)=-\<g,f\>,
\label{skewIP}\\
&&\<R_{2k}, R_{2k+1}\>=-\<R_{2k+1}, R_{2k}\>=r_k,\quad \mbox{all others}=0.
\label{skeworthogonality}
\eea
Under this choice, $F(x,x')$ itself is expressed as
\be
F(x,x')=\sum_{k=0}^{\infty}\Bigl(\phi_{2k}(x) \phi_{2k+1}(x')-\phi_{2k+1}(x) \phi_{2k}(x')\Bigr).
\ee
Thus the matrix element of $J_N
=[J(x_i,x_j)]_{i,j=1,\ldots,N}
=[I(x_i,x_j)+F(x_i,x_j)]_{i,j=1,\ldots,N}$ takes the form
\be
J(x,x')=\sum_{k=N/2}^{\infty}\Bigl(\phi_{2k}(x) \phi_{2k+1}(x')-\phi_{2k+1}(x) \phi_{2k}(x')\Bigr),
\label{Jxx}
\ee
due to the definition (\ref{Ixx}).

One can verify that
\be
R_{2k}^{(0)}(x)=\sum_{j=0}^k
\frac{2^{2k} k! \Gamma\(k+(\nu+1)/2\)}{2^{2j}j! \Gamma\(j+(\nu+1)/2\)!}(2j)! 
L_{2j}^{\nu-1}(x)\ \ \ \mbox{and}\ \ \ 
R_{2k+1}^{(0)}(x)=-(2k+1)! L_{2k+1}^{\nu-1}(x)
\ee
satisfy the skew-orthogonality (\ref{skeworthogonality}) at $\tau=0$ 
(i.e.,~chGSE, with the skew inner product denoted by  $\<\ , \ \>_{(0)}$)
with $r_k^{(0)}=(2k+1)!(2k+\nu)!$ \cite{NW92}.
Due to a lemma \cite{FP95}, 
\be
\bigl< R_m, R_n \bigr>= e^{-(\gamma_m+\gamma_n)\tau} \bigl< R_m^{(0)}, R_n^{(0)} \bigr>_{(0)}
\ee
which directly follows from the definition (\ref{skewIP}),
\be
R_{2k}(x)=\sum_{j=0}^k
\frac{2^{2k} k! \Gamma\(k+(\nu+1)/2\)}{2^{2j}j! \Gamma\(j+(\nu+1)/2\)!}(2j)! 
L_{2j}^{\nu-1}(x) e^{(\gamma_{2j}-\gamma_{2k})\tau}\ \ \ \mbox{and}\ \ \ 
R_{2k+1}(x)=R_{2k+1}^{(0)}(x)
\ee
satisfy the skew-orthogonality (\ref{skeworthogonality}) at $\tau>0$ 
with $r_k=r_k^{(0)} e^{-(\gamma_{2k}+\gamma_{2k+1})\tau}$ \cite{FNH99}.
Together with the definitions (\ref{psi})--(\ref{Sxx}) and (\ref{Jxx}),
the quaternion kernel elements at finite $N$ are completely determined.

\subsection{Microscopic crossover scaling limit}\label{scalinglimit}
Now we concentrate on the case in which the Kramers degeneracy is weakly broken by 
the small parameter $\tau\ll 1$.
Then the spectral density ${\rho}(\lambda)$ of $H$ in the large-$N$ limit
is identical to that of chGSE ($\tau=0$), i.e., Wigner's semicircle
${\rho}(\lambda)=\pi^{-1}\sqrt{4N-\lambda^2}$.
We magnify the vicinity of the origin by introducing the rescaled variables\footnote{%
Here we have abused the notation slightly: 
Up to Sect.~\ref{SOP}, $x_i=\lambda_i^2$ denotes squared eigenvalues of $H$,
whereas after Sect.~\ref{scalinglimit} $x_i=\lambda_i/\varDelta$ denote microscopically rescaled
eigenvalues of $H$.
Accordingly the function symbols $K(x,y)$, $S(x,y)$, etc.~are also used with two different meanings.}
$x_i:=\lambda_i/\varDelta$
which measure the eigenvalues in units of the mean level spacing
at the origin, $\varDelta=1/{\rho}(0)=\pi/\sqrt{4N}$.
Moreover, in order to realize a nontrivial crossover behavior we 
take the triple-scaling limit $N\to\infty, \lambda\to 0, \tau\to 0$
while keeping the combination $\rho=\sqrt{\tau}/\varDelta$ and $x_i$ fixed finite.
In this limit, sums over $k$ turn to integrals over $v:=k/N$ and
Laguerre polynomials reduce to Bessel functions,
$\sqrt{w(z)}L^\nu_k(z)\sim k^{\nu/2} J_\nu(2\sqrt{kz})$ as $k\to\infty$.
Accordingly, the quaternion kernel elements (\ref{Dxx}), (\ref{Sxx}), (\ref{Jxx}) reduce to (see footnote ${}^{\dagger}$) 
\cite{FNH99}
\bea
S(x,y)\!\!\!\!&=&\!\!\!\!
\pi  \sqrt{x y} \left\{
\frac{J_{\nu}(\pi  x) y J_{\nu-1}(\pi  y)-x J_{\nu-1}(\pi  x) J_\nu(\pi  y)}{x^2-y^2}
\right. \nn\\
&&
\left. ~~~~~~~~~~ -\frac{J_\nu(\pi x)}{2}
\pi \int_0^1 dv\, e^{\pi^2\rho^2 (v^2-1)}J_\nu(\pi v y)
\right\} ,
\label{Smicro}\\
D(x,y)\!\!\!\!&=&\!\!\!\!
\frac{\pi^2 \sqrt{xy} }{2} \int_0^1 dv\,v  \int_0^1 du\, e^{\pi^2 \rho^2 v^2(1+u^2)}
\left\{J_\nu(\pi v u x) J_\nu(\pi v y)- J_\nu(\pi v x) J_\nu(\pi v u y)\right\},
\label{Dmicro}\\
J(x,y)\!\!\!\!&=&\!\!\!\!
\frac{\pi^3\sqrt{x y}}{2}  \int_1^\infty dv\, v^2 \, e^{-2 \pi^2\rho^2 v^2} 
\left\{J_{\nu}(\pi v x) y J_{\nu-1}(\pi v y)- x J_{\nu-1}(\pi v x) J_{\nu}(\pi v y)\right\}.
\label{Jmicro}
\end{eqnarray}
Thus the correlation function of $n$ positive rescaled eigenvalues $\{x_i\}$ of $H$
 in the vicinity of the origin 
is finally expressed as 
\bea
R_{n}(x_1,\ldots,x_n)=
\mathrm{Pf} \(Z\left[K(x_i, x_j)\right]_{i,j=1}^n \),
\ \ 
K(x, y)=\left[
\ba{cc}
S(x,y) & J(x, y)\\
D(x,y) & S(y, x)
\ea
\right].
\label{KSDI}
\eea

\subsection{Individual eigenvalue distributions}
It is well known that, for a determinant process in which 
$n$-point correlation functions are expressed in terms of a scalar kernel $K(x,y)$
as 
$R_{n}(x_1,\ldots,x_n)=\det \left[K(x_i, x_j)\right]_{i,j=1}^n$, 
the probability $E_{\ell}(I)$
for an interval $I$ to contain exactly $\ell$ points is expressed as 
a Fredholm determinant (Det) of $K$ over $I$ \cite{Meh04}: 
\be
E_{\ell}(I)={\rm Prob}[\#(I)=\ell]=
\frac{1}{\ell!}\left(-\frac{\partial}{\partial\xi}\right)^\ell
\left.{\rm Det} (\mathbb{I}-\xi \hat{K}_I)\right|_{\xi=1}.
\label{FredholmDet}
\ee
Here $\hat{K}_I$ acts on $L^2$-functions $f(x)$ over $I$ as
$(\hat{K}_I f)(x)=\int_I dy\,K(x,y) f(y)$.
This argument directly carries over to our case of the quaternion determinant process in which
$n$-point correlation functions are expressed in terms of a quaternion kernel $\bs{K}(x,y)$ as 
$R_{n}(x_1,\ldots,x_n)=\mathrm{qdet} \,\bs{K}_n(x_1,\ldots,x_n)=
\mathrm{Pf}\(Z \left[K(x_i, x_j)\right]_{i,j=1}^n\)$, 
leading to
\be
E_{\ell}(I)
=
\frac{1}{\ell!}\left(-\frac{\partial}{\partial\xi}\right)^\ell
\left.{\rm Pf} (Z-\xi Z\hat{K}_I)\right|_{\xi=1}
=
\frac{1}{\ell!}\left(-\frac{\partial}{\partial\xi}\right)^\ell
\left.{\rm Det} (\mathbb{I}-\xi \hat{K}_I)^{1/2}\right|_{\xi=1}.
\label{FredholmPfaffian}
\ee
This time, $\hat{K}_I$  acts on 2-component $L^2$-functions $F(x)$ over $I$ as
$(\hat{K}_I F)(x)=\int_I dy\,K(x,y) \cdot F(y)$.
By differentiating 
${\rm Det} (\mathbb{I}-\xi \hat{K}_I)^{1/2}=\exp \frac12 {\rm Tr} \log(\mathbb{I}-\xi \hat{K}_I)$ in $\xi$, 
the first few $E_\ell(s)$ are expressed 
in terms of the Fredholm determinant and  the resolvents of the operator $\hat{K}_I$,
\[
T_n(I):=\frac12 {\rm Tr} \bigl(\hat{K}_I(\mathbb{I}-\hat{K}_I)^{-1}\bigr)^n,
\]
as \cite{Nis15} 
\begin{eqnarray}
\!\!\!\!\!&&
E_0(I)=
{\rm Det}({\mathbb I}-\hat{K}_I)^{1/2}
\nn
\\
\!\!\!\!\!&&
E_1(I)=E_0 \,T_1
\nn\\
\!\!\!\!\!&&
E_2(I)= \frac{E_0}{2!} \(T_1^2 -T_2 \)
\nn\\
\!\!\!\!\!&&
E_3(I)= \frac{E_0}{3!} \(T_1^3 -3T_1 T_2  +2T_3 \)
\nn\\
\!\!\!\!\!&&
E_4(I)= \frac{E_0}{4!} \(T_1^4 -6 T_1^2 T_2 + 3T_2^2 +8 T_1 T_3 -6 T_4 \) 
\label{Ek}\\
\!\!\!\!\!&&
E_5(I)= \frac{E_0}{5!}
\(
T_1^5 - 10 T_1^3 T_2 + 20 T_1^2 T_3 + 15 T_1 T_2^2
- 30 T_1 T_4 - 20 T_2 T_3 + 24 T_5
\)
\nn\\
\!\!\!\!\!&&
E_6(I)=\frac{E_0}{6!}
\left\{
\begin{array}{l}
T_1^6 - 15T_1^4 T_2 + 40T_1^3 T_3 + 45T_1^2 T_2^2- 90T_1^2 T_4- 120T_1 T_2 T_3 - 15T_2^3 \\
+ 144T_1 T_5+ 90T_2  T_4+ 40T_3^2 -120 T_6
\end{array}
\right\}
\nn\\
\!\!\!\!\!&&
E_7(I)=
\frac{E_0}{7!}
\left\{
\begin{array}{l}
T_1^7 - 21T_1^5 T_2  + 70T_1^4 T_3 + 105T_1^3 T_2^2 - 210 T_1^3 T_4 - 420T_1^2 T_2  T_3
- 105 T_1 T_2^3\\
+ 504 T_1^2 T_5+ 630 T_1 T_2 T_4 + 280 T_1 T_3^2 
 +   210 T_2^2 T_3  - 840 T_1 T_6 -  504 T_2 T_5 \\
 - 420 T_3 T_4 + 720 T_7
 \end{array}
\right\} .
\nn 
\end{eqnarray}
After specializing to $I=[0,s)$ and abbreviating $E_\ell(s):=E_\ell\([0,s)\)$,
the probability distribution $p_k(s)$ of the $k {\rm th}$ smallest positive eigenvalue
is given in terms of $E_{0}(s), \ldots, E_{k-1}(s)$ as
\be
p_k(s)=-\frac{d}{ds}\sum_{\ell=0}^{k-1} E_{\ell}(s).
\label{pks}
\ee
This relationship follows from a simple observation that, 
for a joint of two intervals $[0, s+ds)=[0,s) \cup[s,s+ds):=I \cup dI$,
the probability that the narrower interval $dI$ contains more than one eigenvalue
is of order $O(ds^2)$, so to the order $O(ds^1)$ one has
\begin{equation}
E_{\ell}(s+ds)
\simeq
{\rm Prob}[\#(I)=\ell \cap \#(dI)=0]+
{\rm Prob}[\#(I)=\ell-1 \cap \#(dI)=1].
\label{Elsds}
\end{equation}
Subtracting Eq. (\ref{Elsds}) from the definition of $E_\ell(s)$ gives
\be
E_{\ell}(s)-E_{\ell}(s+ds)\simeq
{\rm Prob}[\#(I)=\ell \cap \#(dI)=1]-{\rm Prob}[\#(I)=\ell-1 \cap \#(dI)=1]
\ee
which is equivalent, in the limit $ds\searrow 0$, to
\begin{equation}
-\frac{d}{ds} E_{\ell}(s) =p_{\ell+1}(s)-p_{\ell}(s) ,
\end{equation}
with $p_{0}(s)= 0$ understood.
Summing over $\ell=0,\ldots,k-1$ gives Eq. (\ref{pks}).

An efficient way of numerically evaluating the Fredholm determinant of
a trace-class operator $\hat{K}_I$
is the Nystr\"{o}m-type discretization \cite{Bor10-1,Bor10-2}
\be
{\rm Det}(\mathbb{I}-\hat{K}_I)\simeq \det(\openone_M-\mathcal{K}_I),\ \ \mbox{where}\ \ 
\mathcal{K}_I= \left[K(x_i,x_j) \sqrt{w_i\,w_j}\right]_{i,j=1}^M
\label{Nystrom}
\ee
is an $M\times M$ matrix evaluated with
a quadrature rule consisting of a set of $M$ points $\{x_i\} \in I$ and associated weights $\{w_i\}$ such that
${\int_I f(x)dx  \simeq \sum_{i=1}^M f(x_i) w_i}$.
As the order $M$ of the quadrature increases, 
the RHS of Eq. (\ref{Nystrom}) is proven to converge to its LHS uniformly and 
exponentially fast in $M$ \cite{Bor10-1,Bor10-2}.
We also need to evaluate the resolvents in Eq. (\ref{Ek}), which are likewise approximated as
\be
{\rm Tr}\bigl( \hat{K}_I(\mathbb{I}-\hat{K}_I)^{-1}\bigr)^n
\simeq {\rm tr} \left(\mathcal{K}_I(\openone_M-\mathcal{K}_I)^{-1}\right)^n.
\ee
Obviously these formulae hold for a $2\times 2$-matrix-valued kernel as well.
For our purpose we employ the Gauss--Legendre quadrature rule in which $\{x_1,\ldots,x_M\}$ are
the nodes of the Legendre polynomial $P_M(x)$ on a shifted and rescaled domain $[-1,1]\mapsto I=[0,s]$.
We have applied the Nystr\"{o}m-type method 
to the kernel (\ref{Smicro})--(\ref{KSDI})
for the chGSE--chGUE crossover at $\nu=0$, bearing in mind that 
the topological charge $\nu$ is washed away in the staggered Dirac operator
that we shall employ in Sect.~3.
We have numerically evaluated $p_1(s), \ldots, p_4(s)$
with $M$ at least 20 for far-more-than-sufficient precision,
and confirmed the stability of the results for increasing $M$.
Plots of $p_1(s), \ldots, p_4(s)$ for $0\leq s\leq 5.5$ are exhibited in Fig.~1 (left).

\begin{figure}[h] 
\begin{center}
\includegraphics[bb=0 0 259 165,width=75mm]{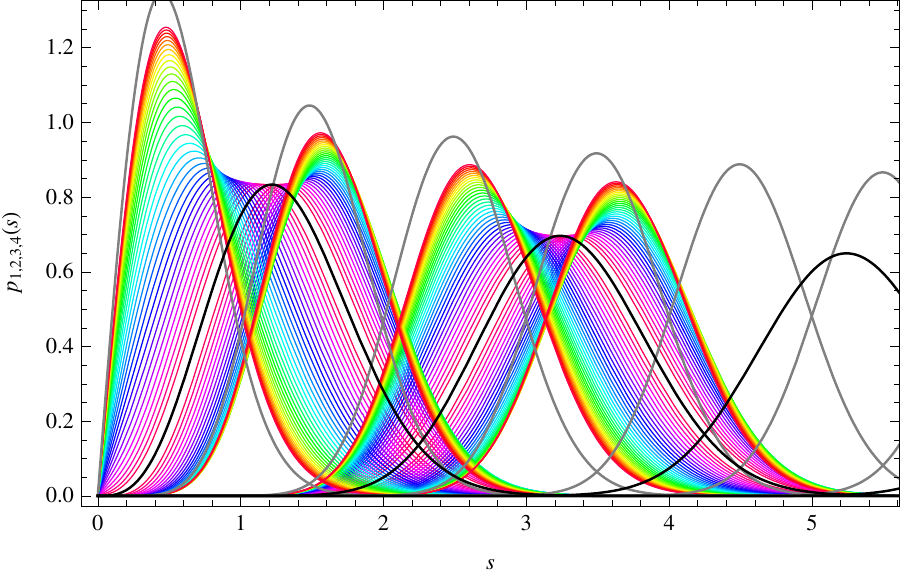}
~
\includegraphics[bb=0 0 258 165,width=74.2mm]{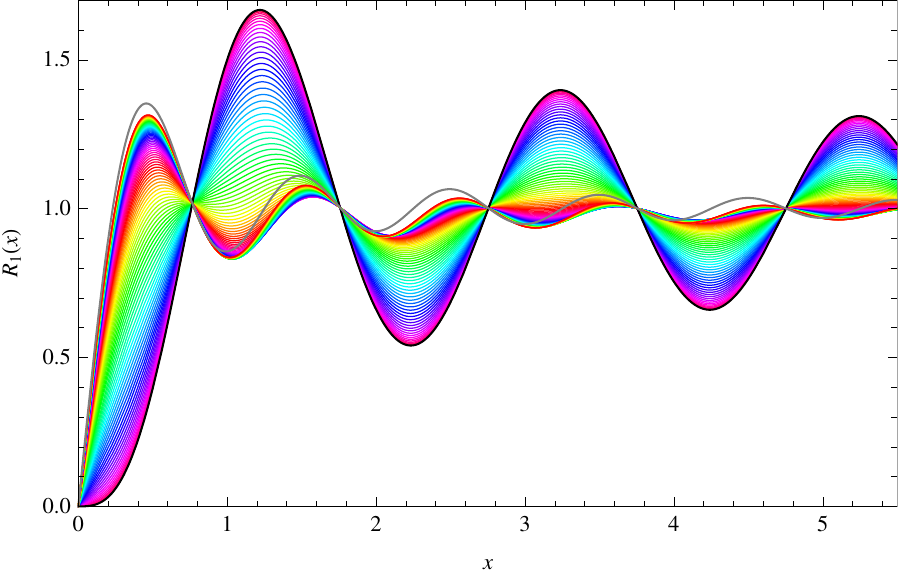}
\caption{
First four eigenvalue distributions $p_1(s), \ldots, p_4(s)$
 (left) for $0.04\leq \rho \leq 0.70$ (step 0.01, purple to red)
and the microscopic spectral density $R_1(x)$ (right) 
for $0.01\leq \rho \leq 1.00$ (step 0.01, purple to red)
for the chGSE (black) to chGUE (gray) crossover, at $\nu=0$.}
\end{center}
\end{figure}

\noindent
For comparison, the spectral densities that comprise the former, 
$R_1(x)=\sum_{k=1}^\infty p_k(x)=S(x,x)$ in Eq. (\ref{Smicro}), are plotted in Fig.~1 (right).
The practical advantage of adopting individual eigenvalue distributions
over $n$-level correlation functions (including $R_1(x)$) for fitting is clear from the figures:
As the oscillation of the latter consists of overlapping multiple peaks,   
the characteristic shape of each peak is inevitably smeared, 
resulting in a rather structureless curve for which an accurate fit is difficult.
On the other hand, the shape of the former is clearly distinguishable and
is extremely sensitive to the $\rho$ parameter, because
the ratio of the two $p_k(s)$ of the chGUE--chGSE crossover at different $\rho$
grows as $\exp (\mathrm{const.}s^2)$ for large $s$.
Therefore, the $p_k(s)$ are expected to admit very precise one-parameter fitting of data by the least-squares method
(as will be shown in Tables 3--6 of Sect.~3).

\subsection{Effective theory and low-energy constants}
Now we shall relate the crossover random matrix ensemble to the nonlinear $\sigma$ models
originating from gauge theory.
In continuum, 
QCD-like theories with $N_F$ flavors of quarks in a real representation
have pseudoreal Dirac operators,
as does 
our case of SU(2) lattice gauge theory with fundamental staggered fermions
(Eq. (\ref{Dstag}) in Sect.~3).
The low-energy effective Lagrangian of these theories
is universally determined by the spontaneous breaking
of the Pauli--G\"{u}rsey extended flavor group $\mathrm{SU}(2N_F)$ down to its
vector subgroup $\mathrm{SO}(N_F)$ and takes the form
\be
\mathcal{L}_{\rm eff}(Q)=\frac{1}{2} F^2\,{\rm tr}\, \partial_\nu Q^\dagger  \partial_\nu Q
-\frac{1}{2} \Sigma m\,{\rm Re}\,{\rm tr}\, \hat{M} Q
\ee
in the leading order of the $p$-expansion.
Here $Q(x)$ is a symmetric SU($2N_F$) matrix-valued Nambu--Goldstone field
(called a nonlinear $\sigma$ model of class AI),
$m$ the degenerate quark mass, and
$\hat{M}=\sigma_1\otimes \openone_{N_F}$.
It contains two phenomenological constants:
$F$ the pseudo-scalar decay constant and $\Sigma=\<\bar{\psi}\psi\>/N_F$ the chiral condensate
(both measured in the chiral and zero-chemical potential limit $m, \mu\to 0$).
The effect of introducing the quark number chemical potential $\mu$ 
to the fundamental theory
is unambiguously
incorporated in $\mathcal{L}_{\rm eff}$ through flavor covariantization
in the symmetric rank-2 tensor representation \cite{KSTVZ00},
\be
\partial_\nu Q \ \mapsto \ 
\nabla_\nu Q =\partial_\nu Q -i\mu \delta_{\nu 0}(\hat{B}Q+Q\hat{B}),
\ee
with $\hat{B}=\sigma_3 \otimes \openone_{N_F}$.
If the theory is in a finite volume $V=L^4$ and the Thouless energy defined as 
$E_c\sim {F^2}/(\Sigma L^2)$ is much larger than $m$ (called the $\varepsilon$-regime), 
the path integral is dominated by the zero mode only and takes the form
\be
Z=\int dQ\,\exp\(
V\mu^2 F^2\,{\rm tr}\, (\hat{B} Q^\dagger\hat{B} Q+\hat{B}\hat{B})
+\frac12 V\Sigma m \,{\rm Re}\,{\rm tr}\, \hat{M}Q
\).
\label{Zchiral}
\ee
The above {\em form} (not the concrete symmetric space on which $Q$ takes values) 
of action is in fact common for all classes of nonlinear $\sigma$ models
(in even or odd dimensions, with Dyson indices $\beta=1,2,4$) 
in which the global symmetry is broken by the chemical potential \cite{DN03}.

On the other hand, 
the characteristic polynomial 
$\< \det(\lambda-H)^{N_F}\>$ of a random matrix $H$ (\ref{H}) 
in the microscopic crossover scaling limit explained in Sect.~\ref{scalinglimit}
can be evaluated by exponentiating the determinant with $N_F$ flavors of Grassmannian vectors and
using a standard technique of Hubbard--Stratonovich transformations (see, e.g., Ref. \cite{Efe97}) 
for matrices $A$ and $B$.
One can easily show that the outcome is the same nonlinear $\sigma$ model as Eq. (\ref{Zchiral})
with parameters replaced by
\be
VF^2\mu^2\to \frac{\pi^2}{2} \rho^2,  \ \ \ 
V\Sigma m\to i \pi x,
\label{correspondence}
\ee
respectively.
Note that the above identification can also be read off from the exponents 
in the quaternion kernel elements (\ref{Smicro})--(\ref{Jmicro}).

\section{Dirac spectrum} \label{Diracspectrum}
In this section we shall fit
the probability distributions analytically derived in the previous section to
the Dirac operator spectra measured from
two types of lattice gauge simulations:
(a) SU(2) gauge theory with the imaginary chemical potential, and 
(b) SU(2) $\times$ U(1) gauge theory.
In either case the pseudoreality of the staggered SU(2) Dirac operator is
weakly violated by the U(1) field,
which is applied as a fixed background \cite{HOV97} or dynamically fluctuating.

\subsection{Simulation details} \label{simulationdetails}

\noindent
(a) $\mathrm{SU}(2)$ gauge theory with the imaginary chemical potential (ICP):
The $\mathrm{SU}(2)$ variables on temporal links
of a hypercubic lattice of size $V = L^4$ 
are multiplied by a constant phase: 
\begin{equation} 
\tilde{U}_{\nu} (x) ={U}_{\nu} (x)\times
\begin{cases}
\e^{i 2 \pi \varphi} & (\nu = 4,\ x_{4} = L-1) \\ 
1 & \text{(else)} 
\end{cases}. 
\end{equation}
The phase $2 \pi \varphi$ can be regarded as the Aharonov--Bohm (AB) flux 
\cite{DM91}
penetrating the temporal circle
and is gauge-equivalent to the imaginary chemical potential 
$\mu=i \mu_{\mathrm{I}} = 2\pi i \varphi / L$, i.e.,~a 
fixed U(1) background $B_{\nu} = (2 \pi \varphi / L) \delta_{\nu,4}$.
We chose antiperiodic/periodic boundary conditions in the temporal/spatial directions, respectively,
and consider a small twisting along the temporal direction ($\varphi \ll 1$).\\

\noindent
(b) $\mathrm{SU}(2) \times \mathrm{U}(1)$ gauge theory:
Following Ref.~\cite{BDHIY07},
non-compact $\mathrm{U}(1)$ link variables $B_{\nu}(x)$
are generated under the Coulomb gauge-fixing condition
(with an additional constraint for the $B_{4}(x)$)
and at the unit coupling constant,
and are multiplied to $\mathrm{SU}(2)$ link variables $U_{\nu}(x)$,
\begin{equation} 
\tilde{U}_{\nu}
 (x) =
U_{\nu}(x)\, e^{i e B_{\nu}(x)}  ,
\end{equation}
with $e$ denoting the bare $\mathrm{U}(1)$ coupling constant.
Fermions are quenched both for the SU(2) and U(1) gauge fields.
For the pure SU(2) case $e=0$ (or $\varphi=0$),
the staggered $\mathrm{SU}(2)$ Dirac operator in the fundamental representation
\beq
D_{x,y}=\sum_{\nu=1}^4 (-1)^{\sum_{i=1}^{\nu-1} x_i} 
\(\tilde{U}_\nu(x)\delta_{x,y+\hat{\nu}}-\tilde{U}^\dagger_\nu(x)\delta_{x,y-\hat{\nu}}\)
\label{Dstag}
\eeq
is pseudoreal,
i.e.,~satisfies
\beq
\mathcal{T} D \mathcal{T}^{-1}=D\ \  \text{with}\ \ 
\mathcal{T}^2=(Z\mathcal{C})^2=-\openone
\label{pseudoreality}
\eeq
($\mathcal{C}$ denotes complex conjugation)
for either choice of
periodic or antiperiodic boundary condition in each direction, and 
we impose periodic boundary conditions in all four directions and
consider the $\mathrm{U}(1)$ part as a small perturbation (i.e.,~$e \ll 1$).\\

As the presence of the AB flux $\varphi$ or the $\mathrm{U}(1)$ coupling $e$ 
breaks the pseudoreality (\ref{pseudoreality}), they
parametrize antiunitary symmetry 
breaking in Dirac operators and are anticipated to be the lattice gauge theory counterpart of
the crossover parameter $\rho$ in random matrix ensemble interpolating chGSE and chGUE. 
We note that while the effect of the ICP $\mu_{\mathrm{I}}$
on the low-energy effective Lagrangian is completely dictated on the symmetry ground \cite{SV05,MT05}
and is related to the pseudo-scalar decay constant $F$ as for the real chemical potential \cite{KSTVZ00}, 
the effect of $e$ cannot be directly related to $F$,
because integrating over the dynamical U(1) gauge field
in the effective Lagrangian would lead to nonlocal self-couplings of pseudo-scalar mesons.

\subsubsection{Simulation setup}
We measured low-lying spectra of the na\"{\i}ve staggered Dirac operator
(\ref{Dstag}) on small lattices of volume $V=4^4$ and $6^4$.
In order to examine the validity of our method for 
the strong-coupling to the near-continuum scaling regions,
we chose the bare $\mathrm{SU}(2)$ gauge coupling constant $\beta=4/g^2$ from the range
$\beta=0, 0.25,\ldots, 1.75$ (step $0.25$) on $V=4^4$ and
$\beta=0, 0.25,\ldots, 2.0$ (step $0.25$), $2.1$ on $V=6^4$.
The simplest algorithm is employed in generating SU(2) gauge configurations: 
unimproved plaquette action and the $10$-hit heat-bath update combined with over-relaxation. 
The antiunitary symmetry violation parameters are set to be:
(a) 
$\varphi=0.01, \ldots, 0.06$ (step 0.01) on $V=4^4$ and 
$\varphi=0.01, \ldots, 0.05$ (step 0.005) on  $V=6^4$, and
(b) 
$e=0.002,\ldots,  0.006$ (step $0.001$),  0.008, 0.0010 on $V=4^4$ and 
$e=0.0004, \ldots, 0.0016$ (step $0.0002$), 0.0020, 0.0024, 0.0028 on $V=6^4$.
$N_{\text{conf}} = 40000$ (10000) configurations are generated and diagonalized on 
$V=4^4$ $(6^4)$ for each set of parameters $(\beta, \varphi)$ or $(\beta, e)$.

\subsubsection{Fitting Dirac spectra}
Our procedure of 
fitting the Dirac spectra to the
individual eigenvalue distributions of
the crossover chiral random matrices 
consists of the following two steps: 
\begin{enumerate}
\item \textit{Determination of the mean level spacing ${\varDelta}$ at the origin}. ~ 
In order to determine the physical scale of Dirac eigenvalues
upon which the effects of perturbations are to be evaluated,
we measure for $N_{\text{conf}}$ independent configurations
four low-lying nondegenerate eigenvalues
$\lambda_{2i-1}=\lambda_{2i}\ (i=1,\ldots,4)$ 
of the pure SU(2) Dirac operator
(due to its quaternionic nature, all eigenvalues are doubly degenerate).

For each $i$,
the mean level spacing $\varDelta_{i}$ at the spectral origin 
is determined by best-fitting the histogram of the unfolded Dirac eigenvalue
$\lambda_{2i}/\varDelta_{i}$ to 
the normalized individual eigenvalue distribution $p_{i}(s)$ of chGSE
so that $\chi^2/$d.o.f.~is minimized by varying $\varDelta_{i}$.
In doing so, we discard the left tail $[0,s_{\mathrm{min}}]$ and the right tail $[s_{\mathrm{max}},\infty)$
of the probability distribution $p_{i}(s)$ for which 
$\int_0^{s_{\mathrm{min}}}ds\,p_i(s)=\int_{s_{\mathrm{max}}}^\infty ds\,p_i(s)\simeq 10^{-3}$,
and split the mid-range into
$B$ bins of fixed widths $\delta s = 0.1$, 
$[s_{\mathrm{min}},s_{\mathrm{max}}]=I_1\cup\cdots\cup I_B$.
Then we define $\chi^2$ from the measured frequency $F_{b}=\#\{\lambda_{2i}\in I_b\}$
and its analytic prediction $f_{b}= N_{\text{conf}} \int_{I_{b}} ds\,p_{i}(s)$ by
$
\chi^{2} = \sum_{b=1}^{B} (F_{b} - f_{b})^2/f_{b}.
$
The statistical error $\delta\varDelta_{i}$ is estimated as a deviation from the optimal $\varDelta_{i}$ 
at which $\chi^2/$d.o.f.~increases by unity.
The combined value of the mean level spacing at the origin $\bar{\varDelta}$ is obtained
as the weighted average of $(\varDelta_{i}, \delta\varDelta_{i}),\ i=1,\ldots,4$.
We have confirmed that these four data are always mutually consistent, so that their combination
helps to improve the statistical error in $\bar{\varDelta}$
as compared to the previous method of using the smallest eigenvalue only \cite{Nis12-1,Nis12-2}.

\item \textit{Determination of the crossover parameter $\rho$}. ~ 
Next we switch on the U(1) perturbation (a) or (b)
and measure the Dirac spectra  $\{ \lambda_{k} \}$
for $N_{\text{conf}} = \mathcal{O} (10^4)$ independent configurations. 
The effect of such perturbations on $\varDelta$ (i.e.,~the chiral condensate) is 
negligible in the lowest order 
in the $\varepsilon$-expansion that we are working on.
U(1) perturbations split once-Kramers-degenerate pairs of eigenvalues,
$\lambda_{2i-1} <\lambda_{2i}$.
We take the first two pairs of Dirac eigenvalues $(\lambda_{2i-1}, \lambda_{2i}), i=1, 2$,
and define the unfolded eigenvalues as 
$(s_{2i-1}, s_{2i}) = (\lambda_{2i-1}  / \varDelta_{i},  \lambda_{2i} /\varDelta_{i}$).
Then by using the same strategy as in Step (i),
their histograms $P_{k} (s_{k})$ are fitted to 
the analytic results 
$p_{k}(s)$ 
(\ref{Ek}), (\ref{pks}) of the crossover random matrices,
with the crossover parameter $\rho_{k}$ being varied.
The statistical error $\delta\rho_{k}$ of the crossover parameter is again estimated as a deviation
from the optimal $\rho_{k}$ at which $\chi^{2}/\mathrm{d.o.f.}$ increases by unity.

The crossover parameter $\bar{\rho}$ 
corresponding to a particular choice of $\varphi$ or $e$ is eventually
determined as the weighted average of $(\rho_k, \delta \rho_k), \ k=1,\ldots,4$.
The advantage of using both once-degenerate eigenvalue pairs 
(over using, e.g., three low-lying eigenvalues)
is now evident:  Histograms of $s_{2i-1}$ and $s_{2i}$
always shift in the opposite directions under the Kramers-breaking perturbation, so that
the effect of small errors in the determination of the overall scale $\varDelta_i$ in Step (i) (which shifts both histograms
in the same direction) is expected to be canceled in the final value of $\bar{\rho}$ 
obtained by combining $\rho_{2i-1}$ and $\rho_{2i}$.
\end{enumerate}

\subsubsection{Low-energy constants}
Due to the correspondence (\ref{correspondence}),
two low-energy constants contained in the chiral Lagrangian,
the chiral condensate $\Sigma$ and the pseudo-scalar decay constant $F$,
are directly related to $\varDelta$ and $\rho$ measured
in Steps (i) and (ii), 
the former by the Banks--Casher relation 
\begin{equation} 
\Sigma = \frac{\pi }{\varDelta V} ; 
\label{chiralcondensate} 
\end{equation}
the latter for (a) the SU(2)+ICP model by 
\begin{equation} 
F^2 = \frac{\pi^2}{2 V} \left( \frac{{\rho}}{\mu_{\mathrm{I}}} \right)^2 
      = \frac{\pi^2}{2 V} \left( \frac{{\rho}}{2 \pi \varphi / L} \right)^2 ,
\label{decayconstantSU2ABF}
\end{equation} 
and for (b) the SU(2)$\times$U(1) model by 
\begin{equation} 
\frac{F^2 \mu_{\mathrm{I}}^2}{e^2} = \frac{\pi^2}{2 V}  \left( \frac{\rho}{e} \right)^2.
\label{decayconstantSU2U1}
\end{equation}
Note that the combination ``$F^2 \mu_{\mathrm{I}}^2$"
in the LHS of Eq.~(\ref{decayconstantSU2U1})
is to be regarded as a single coefficient of the $-\tr \hat{B}Q^\dagger\hat{B}Q$ term
in the chiral Lagrangian (\ref{Zchiral}).
As this quantity should be proportional to $e^2$, our aim here is to determine the 
unknown proportionality constant set by the dynamics.
Thus we shall check,  for each $\beta$ in either case of (a) or (b),
the stability of the ratio $R(\varphi)=\rho/\varphi$ or  $R(e)=\rho/e$ as $\varphi$ or $e$ is varied, and determine
its mean value as a weighted average of $\{R(\varphi), \delta R(\varphi)\}$ or
 $\{R(e), \delta R(e)\}$.

\subsection{Simulation results} \label{simulationresults}

\subsubsection{Fitting Dirac spectra}
In Tables 
\ref{tab:SU2ABF_mls} and
\ref{tab:SU2U1_mls}
we exhibit 
optimal values of the mean level spacings $\varDelta_{i}$ of SU(2) Dirac spectra
(a) under the antiperiodic boundary condition on the temporal direction
(to be used for $\mathrm{SU}(2) +$ICP), and
(b) under the periodic boundary conditions on all
directions (to be used for $\mathrm{SU}(2) \times \mathrm{U}(1)$), respectively.
At each $\beta$, we adopt only the $\varDelta_{i}$ 
that pass the $\chi^2$ test of the fitting: $\chi^2/\text{d.o.f.}<2.0$.
In Fig.~2 (top) we exhibit sample plots
of histograms of the four smallest nondegenerate Dirac eigenvalues $P_{i}(\lambda_i)$
versus the corresponding individual eigenvalue distributions $p_i(s)$ of chGSE, 
each being optimally rescaled by $\varDelta_{i}$.
\begin{figure}[b] 
\begin{center}
\includegraphics[width=7.5cm,bb=0 0 260 168]{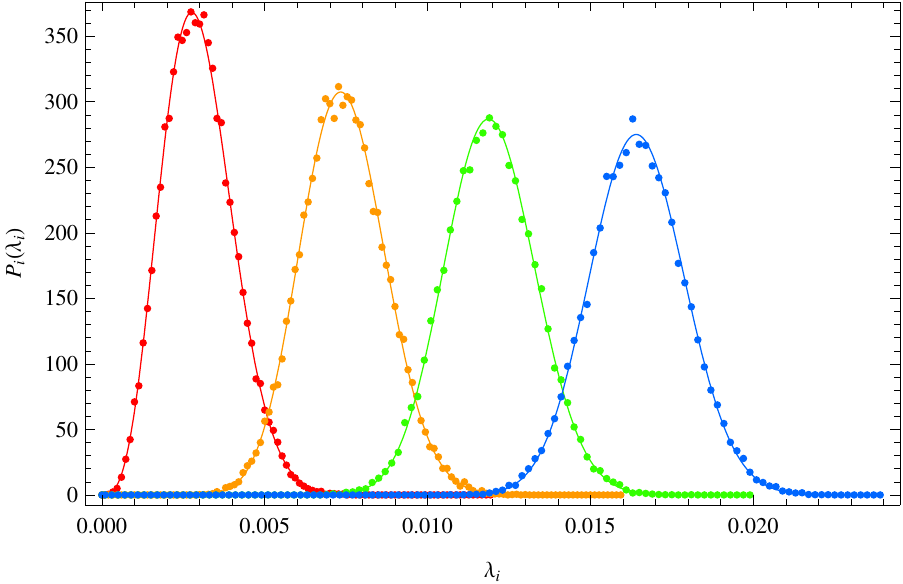}
~
\includegraphics[width=7.5cm,bb=0 0 260 167]{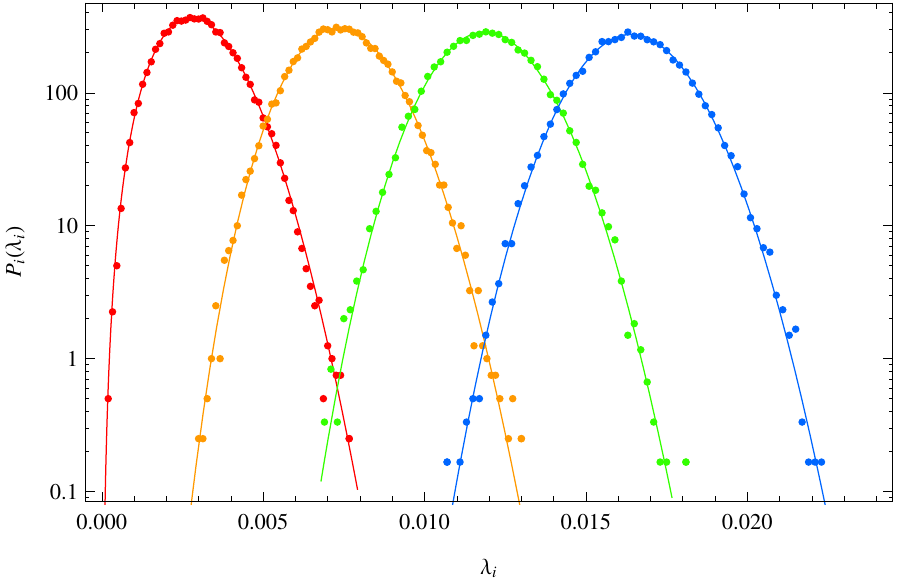}\\
\ \\
\includegraphics[width=7cm,bb=0 0 170 110]{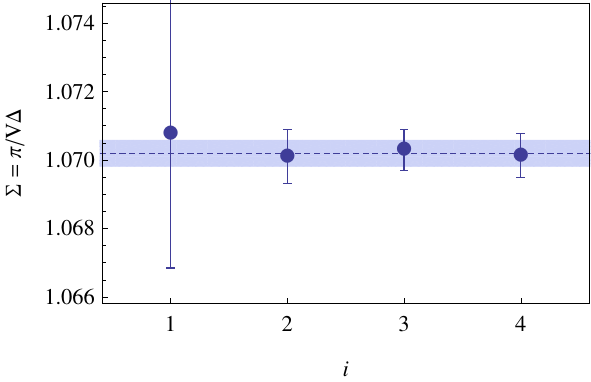}
\caption{
(Top)
Linear and logarithmic plots of
histograms of the four smallest nondegenerate Dirac eigenvalues
$P_{i}(\lambda_i)\ (i=1,\ldots,4)$
(red to blue) of pure SU(2) gauge theory
at $\beta=1.0$, on $V=6^4$ and $N_{\mathrm{conf}}=30000$, and 
individual eigenvalue distributions $p_{i}(s)$ of chGSE,
each being optimally rescaled by a constant $\varDelta_{i}$.
(Bottom)
Values of chiral condensate $\Sigma=\pi/(V\varDelta_{i})$ (circles),
their weighted average (horizontal line), and the combined error (strip).}
\end{center}
\end{figure}
The mutual consistency of such $\varDelta_{i}$ observed in Fig.~2 (bottom) 
justifies the use of the weighted average, 
listed in the seventh columns in Tables \ref{tab:SU2ABF_mls} and \ref{tab:SU2U1_mls}. 
This sample figure illustrates that
the error in $\varDelta_1$ (i.e.,~in $\Sigma$), 
which is relatively larger than that in the other three $\varDelta_{i}$,
is considerably improved
by a factor of 5--10 and is down to $\mathcal{O}(10^{-4})$
by the use of the weighted average of the four.
Note that histograms at $\beta=1.75$ on $V=4^4$ and at $\beta = 2.1$ on $V=6^4$ 
(marked by $*$ in Tables \ref{tab:SU2ABF_mls} and \ref{tab:SU2U1_mls})
failed to be fitted into the chGSE predictions in the above criterion
as $\chi^2 / \text{d.o.f.}$ exceeds 3.
Thus we chose to relax it by cutting off 
the tails of the distributions for which $p_{1} (s) < 0.2$ from fitting,
which leads safely to $\chi^2 / \text{d.o.f.}<2.0$.

\begin{table}[H] 
\caption{
Mean level spacings of the pure $\mathrm{SU}(2)$
Dirac spectrum on $V=4^4$ and $6^4$ with the antiperiodic boundary condition on the temporal direction,
in units of $10^{-2}a^{-1}$.
}
\label{tab:SU2ABF_mls} 
\centering 
\begin{tabular}{lllllllc} \hline \hline
$V$ & $\beta$ & \hspace{15pt}$\varDelta_{1}$ & \hspace{15pt}$\varDelta_{2}$ & \hspace{15pt}$\varDelta_{3}$ & \hspace{15pt}$\varDelta_{4}$ & \hspace{15pt}$\bar{\varDelta}$ & $\chi^2/\text{d.o.f.}$ \\  
\hline 
$4^4$ & 0       & 0.929(2)   & 0.931\hspace{1.5pt}5(7)  & 0.930\hspace{1.5pt}9(5)   & 0.931\hspace{1.5pt}3(4)   & 0.931\hspace{1.5pt}1(3)      & 0.99--1.00  \\ 
      &    0.25   & 0.969(2)   & 0.970\hspace{1.5pt}8(9)  & 0.970\hspace{1.5pt}9(6)   & 0.970\hspace{1.5pt}3(4)   & 0.970\hspace{1.5pt}5(3)      & 0.85--1.00    \\ 
      &    0.5     & 1.019(2) & 1.019\hspace{1.5pt}0(9) & 1.018\hspace{1.5pt}0(6) & 1.017\hspace{1.5pt}5(4) & 1.017\hspace{1.5pt}9(3)     & 0.74--1.02    \\ 
      &    0.75   & 1.073(2) & 1.073\hspace{1.5pt}5(8) & 1.074\hspace{1.5pt}9(6) & 1.075\hspace{1.5pt}6(5) & 1.075\hspace{1.5pt}0(3)      & 1.01--1.36     \\ 
      &    1.0     & 1.151(2) & 1.150\hspace{1.5pt}0(9) & 1.150\hspace{1.5pt}4(7) & --            & 1.150\hspace{1.5pt}3(5)     & 0.68--1.47   \\ 
      &    1.25   & 1.255(2) &  1.254(1)  & --            &  --           & 1.254\hspace{1.5pt}4(9)     & 0.81--1.46    \\ 
      &    1.5    & 1.408(3) & --             & --            & --             & 1.408(3)     & 0.73      \\
      &    1.75$^*$    & 1.705(4) & --            & --            & --            & 1.705(4)      & 0.91   \\   
\hline \hline
$6^4$ & 0       & 0.185\hspace{1.5pt}4(7) & 0.185\hspace{1.5pt}2(3) & 0.185\hspace{1.5pt}0(2) & 0.185\hspace{1.5pt}2(2) & 0.185\hspace{1.5pt}2(1)           &  0.92--1.02   \\ 
      &    0.25   & 0.191\hspace{1.5pt}2(7) & 0.192\hspace{1.5pt}6(3) & 0.193\hspace{1.5pt}1(2) & 0.192\hspace{1.5pt}7(2) & 0.192\hspace{1.5pt}8(1)            & 0.98--1.20   \\ 
      &    0.5     & 0.203\hspace{1.5pt}0(8) & 0.202\hspace{1.5pt}7(4) & 0.202\hspace{1.5pt}0(2) & 0.202\hspace{1.5pt}3(2) & 0.202\hspace{1.5pt}3(1)           & 0.95--0.99    \\ 
      &    0.75   & 0.213\hspace{1.5pt}5(7) & 0.213\hspace{1.5pt}3(2) & 0.214\hspace{1.5pt}3(2) & 0.213\hspace{1.5pt}5(2) & 0.213\hspace{1.5pt}7(1)           & 0.85--1.01     \\ 
      &    1.0     & 0.229\hspace{1.5pt}3(8) & 0.227\hspace{1.5pt}7(4) & 0.227\hspace{1.5pt}9(3) & 0.227\hspace{1.5pt}8(2) & 0.227\hspace{1.5pt}8(1)           & 0.93--1.00    \\ 
      &    1.25   & 0.247\hspace{1.5pt}5(9) & 0.248\hspace{1.5pt}0(4) & 0.247\hspace{1.5pt}9(3) & 0.248\hspace{1.5pt}1(2) & 0.248\hspace{1.5pt}0(2)           & 0.63--0.99     \\ 
      &    1.5     & 0.279\hspace{1.5pt}5(9) & 0.278\hspace{1.5pt}7(5) & 0.278\hspace{1.5pt}3(3) & 0.277\hspace{1.5pt}9(2) & 0.278\hspace{1.5pt}2(2)           & 0.96--1.00     \\ 
      &    1.75   & 0.334(1)  & 0.333\hspace{1.5pt}1(6) & 0.332\hspace{1.5pt}9(4) & --          & 0.333\hspace{1.5pt}0(3)            & 0.73--1.00   \\ 
      &    2.0     & 0.482(2)  & --         & --          & --           & 0.482(2)             & 1.12  \\ 
      &    2.1$^*$    & 0.640(2)  &  --        &  --         &  --          & 0.640(2)             & 1.49   \\ 
\hline \hline 
\end{tabular}
\caption{ %
Mean level spacings of the pure $\mathrm{SU}(2)$
Dirac spectrum on $V=4^4$ and $6^4$ with periodic boundary conditions on all four directions,
in units of $10^{-2}a^{-1}$.
} %
\label{tab:SU2U1_mls} 
\centering 
\begin{tabular}{lllllllc} \hline \hline 
$V$ & $\beta$ & \hspace{15pt}$\varDelta_{1}$ & \hspace{15pt}$\varDelta_{2}$ & \hspace{15pt}$\varDelta_{3}$ & \hspace{15pt}$\varDelta_{4}$ & \hspace{15pt}$\bar{\varDelta}$ & $\chi^2/\text{d.o.f.}$ \\  
\hline 
$4^4$ & 0       & 0.932(2) & 0.930\hspace{1.5pt}6(8) & 0.931\hspace{1.5pt}1(5) & 0.931\hspace{1.5pt}3(4) & 0.931\hspace{1.5pt}1(3)           & 0.62--1.13   \\ 
      &    0.25   & 0.972(2) & 0.971\hspace{1.5pt}8(8) & 0.970\hspace{1.5pt}8(6) & 0.970\hspace{1.5pt}4(4) & 0.970\hspace{1.5pt}8(3)           & 0.84--1.13   \\ 
      &    0.5     & 1.020(2) & 1.016\hspace{1.5pt}0(9) & 1.016\hspace{1.5pt}2(6) & 1.017\hspace{1.5pt}0(4) & 1.016\hspace{1.5pt}7(3)   & 0.97--1.20   \\ 
      &    0.75   & 1.080(2) & 1.077\hspace{1.5pt}0(9) & 1.077\hspace{1.5pt}4(6) & 1.076\hspace{1.5pt}5(5) & 1.076\hspace{1.5pt}9(3)   & 0.76--1.15    \\ 
      &    1.0     & 1.153(2) & 1.152(1) & 1.152\hspace{1.5pt}5(7) & --           & 1.152\hspace{1.5pt}4(5)     & 1.00--1.14    \\ 
      &    1.25   & 1.249(2) & 1.253(1)  & --           &  --          & 1.252\hspace{1.5pt}5(9)     & 0.99--1.00     \\ 
      &    1.5    & 1.412(2) & --                & --        & --           & 1.412(2)     & 0.93    \\ 
      &    1.75$^*$   & 1.698(3) & --               & --        & --           & 1.698(3)     & 1.35  \\ 
\hline \hline
$6^4$ & 0       & 0.186\hspace{1.5pt}5(6) & 0.185\hspace{1.5pt}4(3) & 0.185\hspace{1.5pt}3(2) & 0.185\hspace{1.5pt}4(2) & 0.185\hspace{1.5pt}4(1)   & 0.79--1.00  \\ 
      &    0.25   & 0.192\hspace{1.5pt}8(7) & 0.192\hspace{1.5pt}9(3) & 0.192\hspace{1.5pt}8(2) & 0.192\hspace{1.5pt}8(2) & 0.192\hspace{1.5pt}8(1)   & 0.99--1.00   \\ 
      &    0.5     & 0.202\hspace{1.5pt}4(8) & 0.202\hspace{1.5pt}4(4) & 0.202\hspace{1.5pt}0(2) & 0.202\hspace{1.5pt}0(2) & 0.202\hspace{1.5pt}1(1)   & 0.77--1.01 \\ 
      &    0.75   & 0.214\hspace{1.5pt}6(8) & 0.213\hspace{1.5pt}7(4) & 0.213\hspace{1.5pt}8(2) & 0.213\hspace{1.5pt}6(2) & 0.213\hspace{1.5pt}7(1)   & 0.98--1.01  \\ 
      &    1.0     & 0.228\hspace{1.5pt}6(8) & 0.229\hspace{1.5pt}2(4) & 0.228\hspace{1.5pt}5(3) & 0.228\hspace{1.5pt}5(2) & 0.228\hspace{1.5pt}6(1)   & 0.99--1.00 \\ 
      &    1.25   & 0.248\hspace{1.5pt}1(9) & 0.247\hspace{1.5pt}8(6) & 0.248\hspace{1.5pt}4(3) & 0.247\hspace{1.5pt}7(2) & 0.248\hspace{1.5pt}0(2)   & 0.97--1.01   \\ 
      &    1.5     & 0.277(1) & 0.277\hspace{1.5pt}8(5) & 0.278\hspace{1.5pt}3(3) & 0.278\hspace{1.5pt}2(2) & 0.278\hspace{1.5pt}1(2)     & 0.99--1.00 \\ 
      &    1.75   & 0.333(1)  & 0.331\hspace{1.5pt}8(6) & 0.333\hspace{1.5pt}1(4) & --         & 0.332\hspace{1.5pt}7(3)     & 0.96--1.24 \\ 
      &    2.0     & 0.480(2)  & --         & --          & --          & 0.480(2)      & 1.66 \\ 
      &    2.1$^*$   & 0.639(3) &  --        &  --         &  --         & 0.639(3)      & 0.99 \\ 
\hline \hline 
\end{tabular}
\end{table}%

Next we exhibit the optimal values of the crossover parameters $\rho_{k} (k=1,\ldots,4)$,
determined by fitting individual Dirac eigenvalue histograms to the chGSE--chGUE predictions,
(a) for the SU(2)+ICP model in Tables \ref{tab:SU2ABF_n4_rhok} and \ref{tab:SU2ABF_n6_rhok}, 
and  (b) for the SU(2)$\times$U(1) model in Tables \ref{tab:SU2U1_n4_rhok} and \ref{tab:SU2U1_n6_rhok}. 
Again we adopted only the cases that passed the $\chi^2$ test with $\chi^2/\text{d.o.f.}<2.0$,
and the 
exceptional treatment of 
cutting off the tails of the
smallest eigenvalue distributions for which $p_{1} (s) < 0.2$ from fitting is applied to
$\beta=1.75$ on $V=4^4$ and $\beta=2.1$ on $V=6^4$ (marked by $*$ in Tables 3--6).
Whenever the histogram of the unfolded eigenvalues $s_{2i-1}=s_{2i}$ 
of the pure SU(2) Dirac operator is well fitted into the chGSE prediction in our criterion,
so is the corresponding pair $\{P_{2i-1}(s_{2i-1}), P_{2i}(s_{2i})\}$
of the perturbed Dirac operator to the chGSE--chGUE crossover prediction, as expected.

\begin{figure}[H]  
\centering 
\includegraphics[width=7.5cm,bb=0 0 360 233]{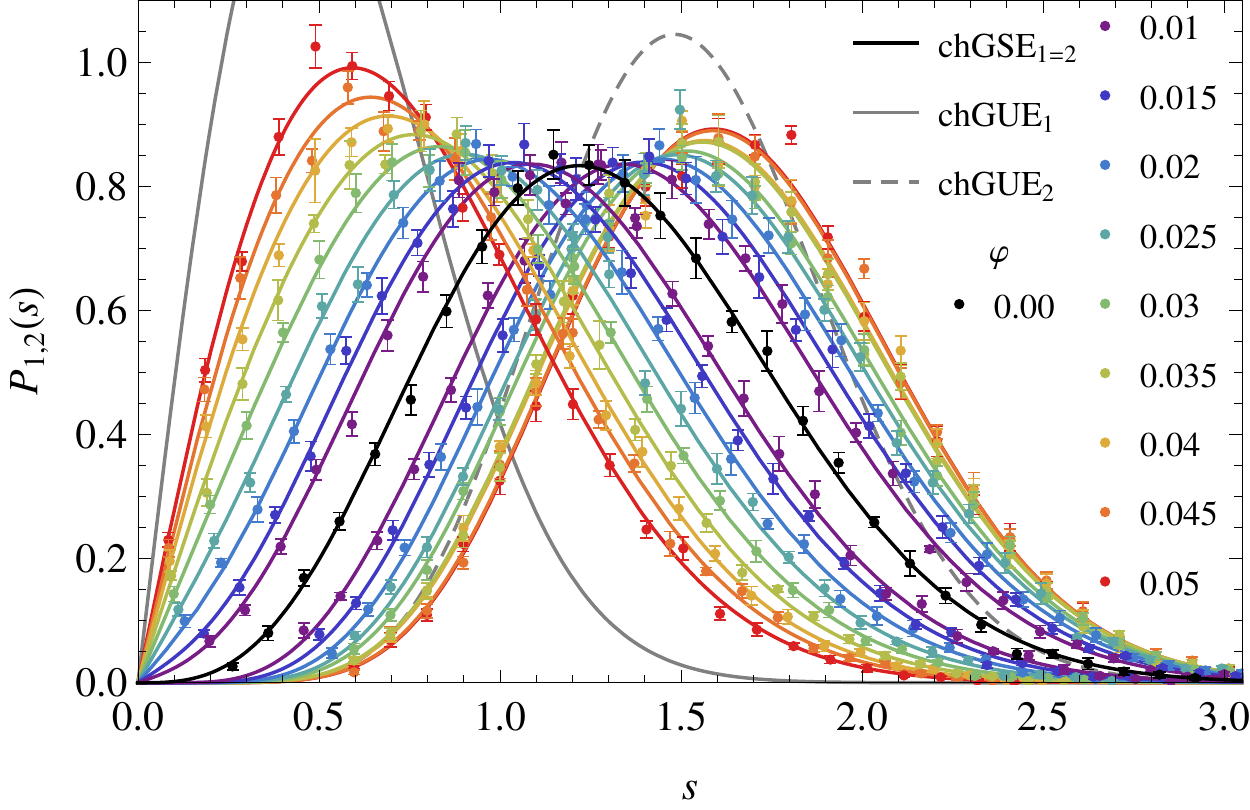}
\includegraphics[width=7.5cm,bb=0 0 360 240]{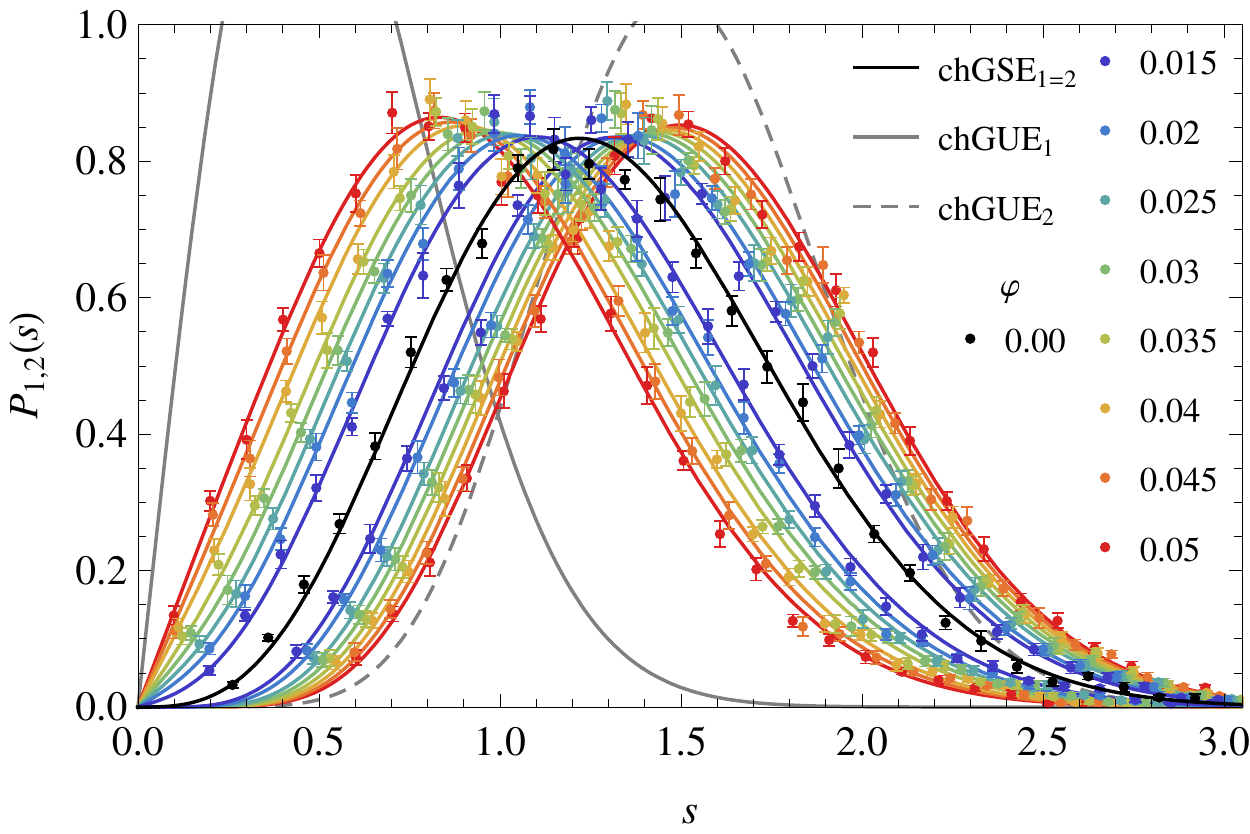}
\includegraphics[width=7.5cm,bb=0 0 360 233]{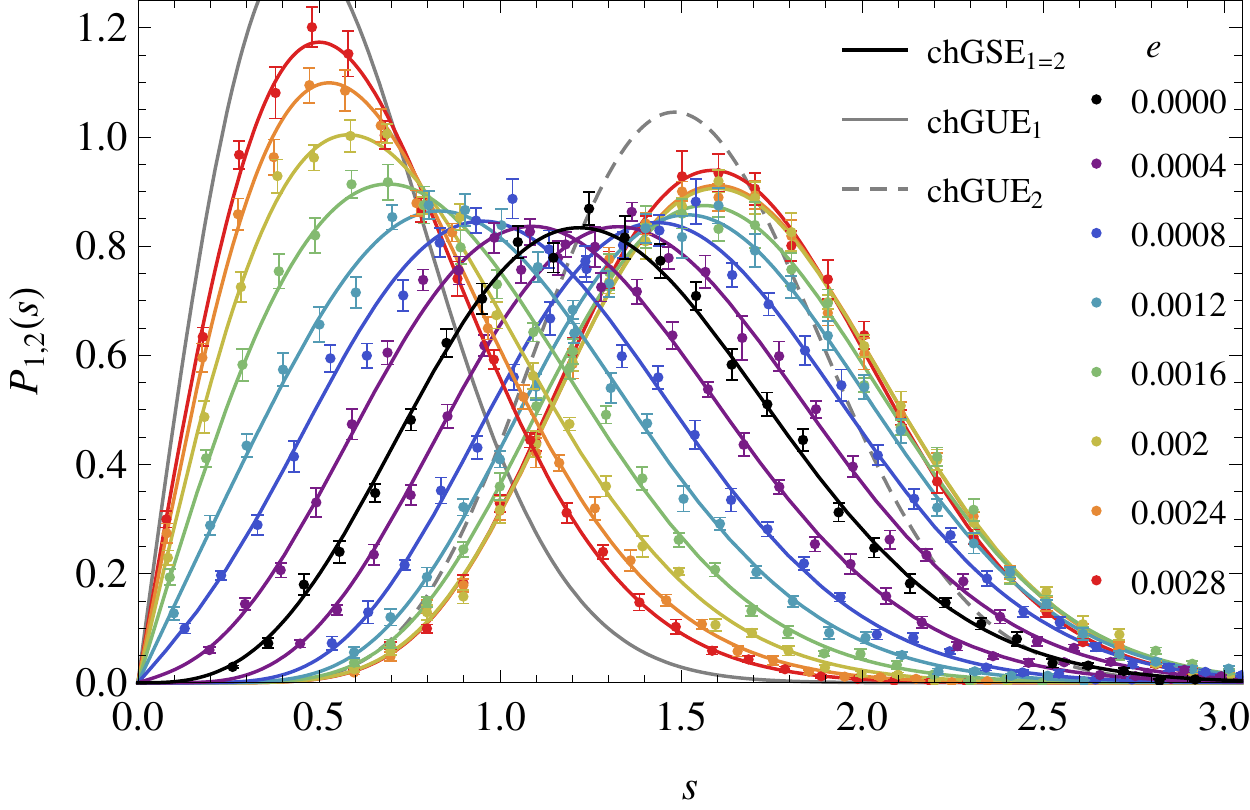}
\includegraphics[width=7.5cm,bb=0 0 360 240]{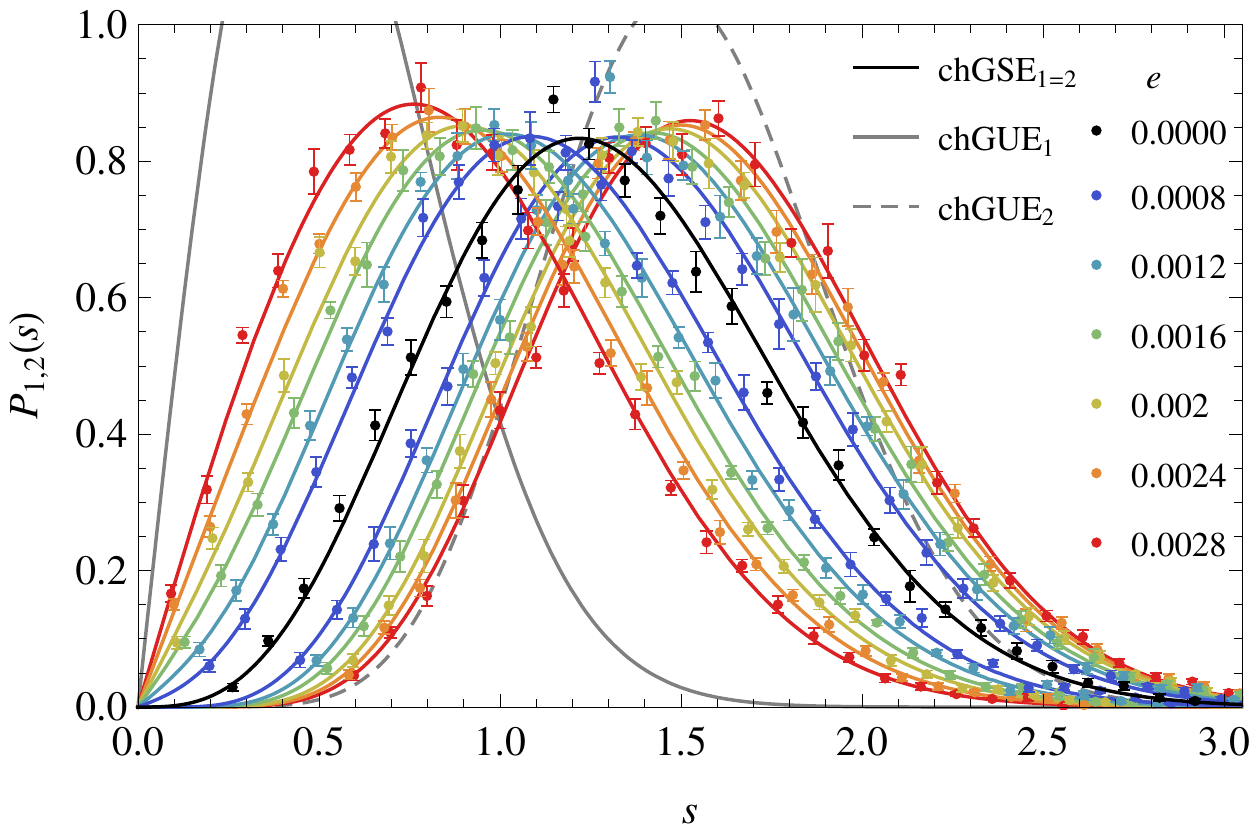}
\caption{%
The first two Dirac eigenvalue distributions $P_{1,2} (s)$ of the 
SU(2)+ICP model (top) and 
SU(2)$\times$U(1) model (bottom) with increasing perturbations $\varphi$ and $e$,
from black ($\varphi, e=0$) to purple (smallest $\varphi, e$) to red (largest $\varphi, e$),
on $V=6^4$ and at $\beta = 1.0$ (left) and at $\beta = 2.0$ (right). 
Those of chGSE and chGUE are plotted in black and gray, respectively,
and the best-fit curves of the chGSE--chGUE crossover are plotted in the same colors as the corresponding
lattice data.
The error bars of the histograms in Figs. 3--5
are estimated by the jackknife method.
}%
\label{fig:SU2U1-SU2ABF_b10-20_p1-2}
\end{figure}

\begin{figure}[H] 
\centering 
\includegraphics[width=7.5cm,bb=0 0 360 234]{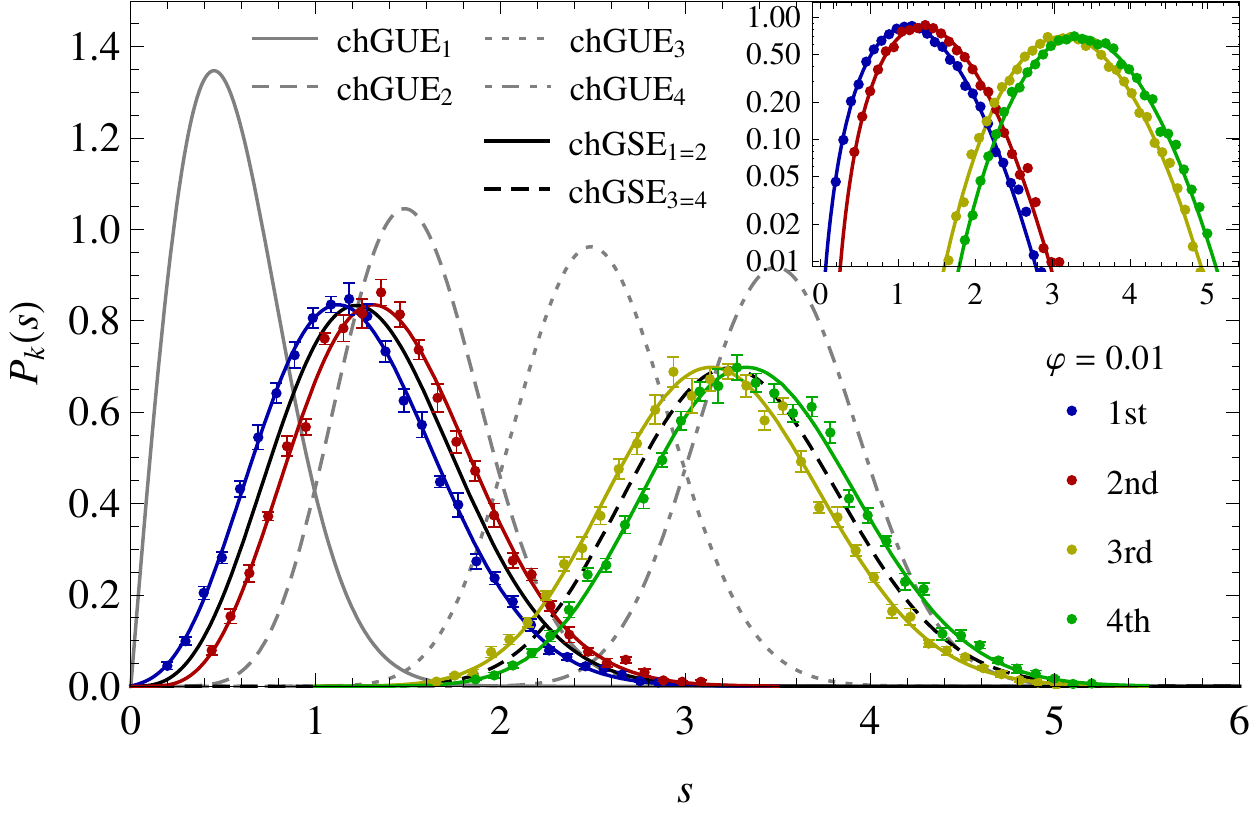}
\includegraphics[width=7.5cm,bb=0 0 360 234]{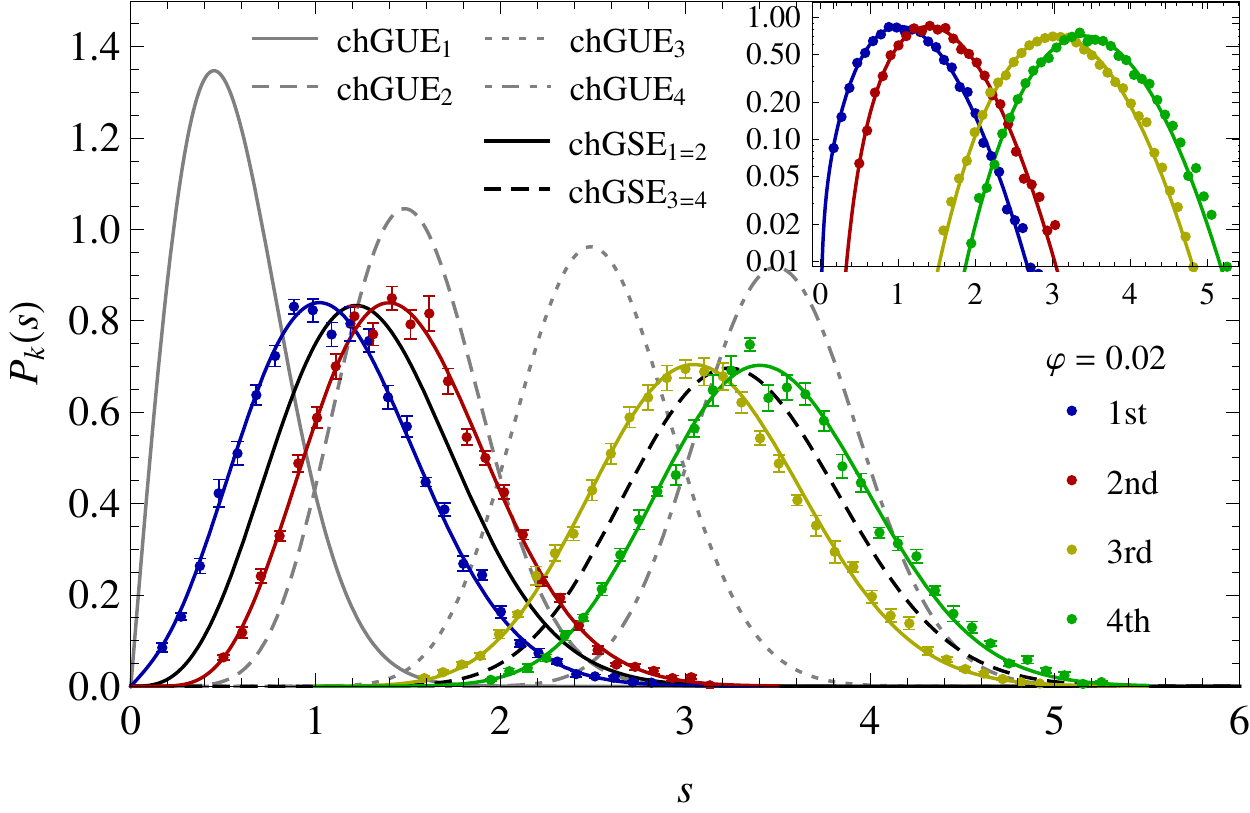}
\includegraphics[width=7.5cm,bb=0 0 360 234]{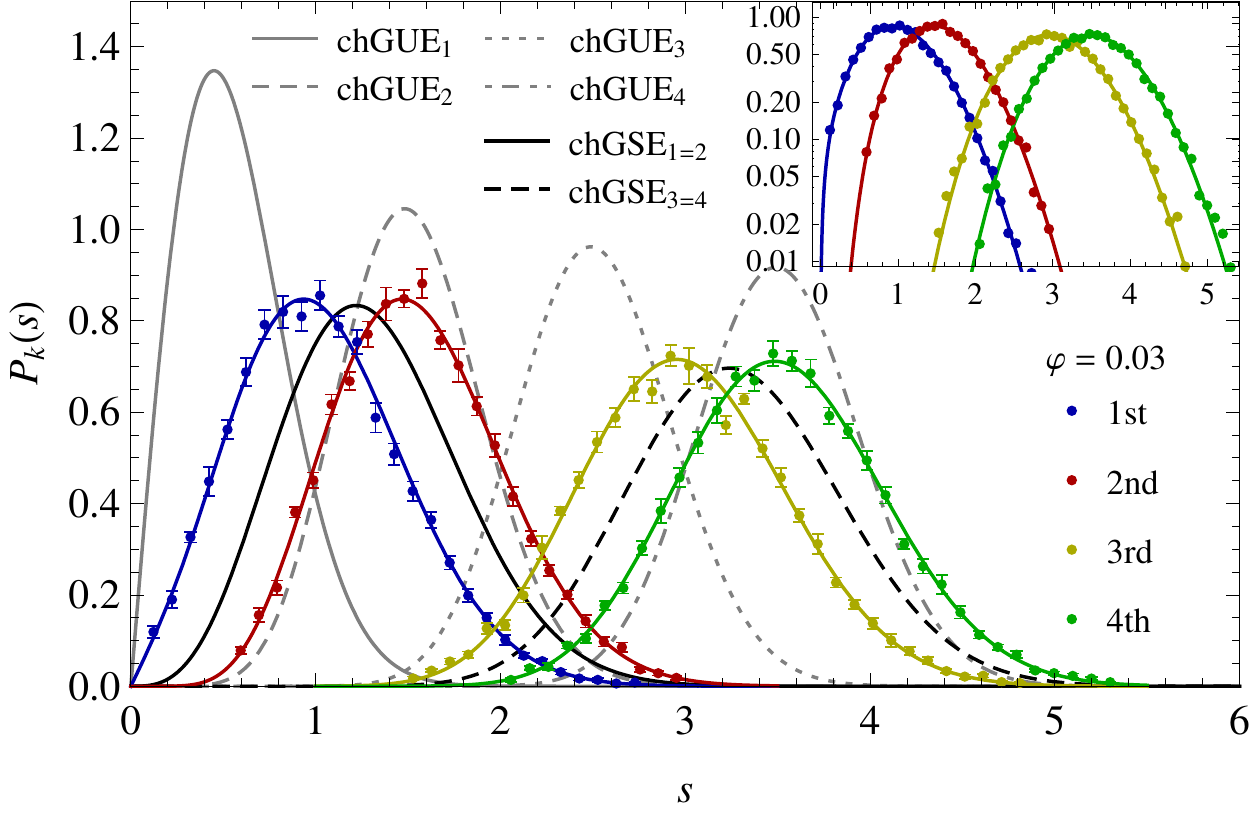}
\includegraphics[width=7.5cm,bb=0 0 360 234]{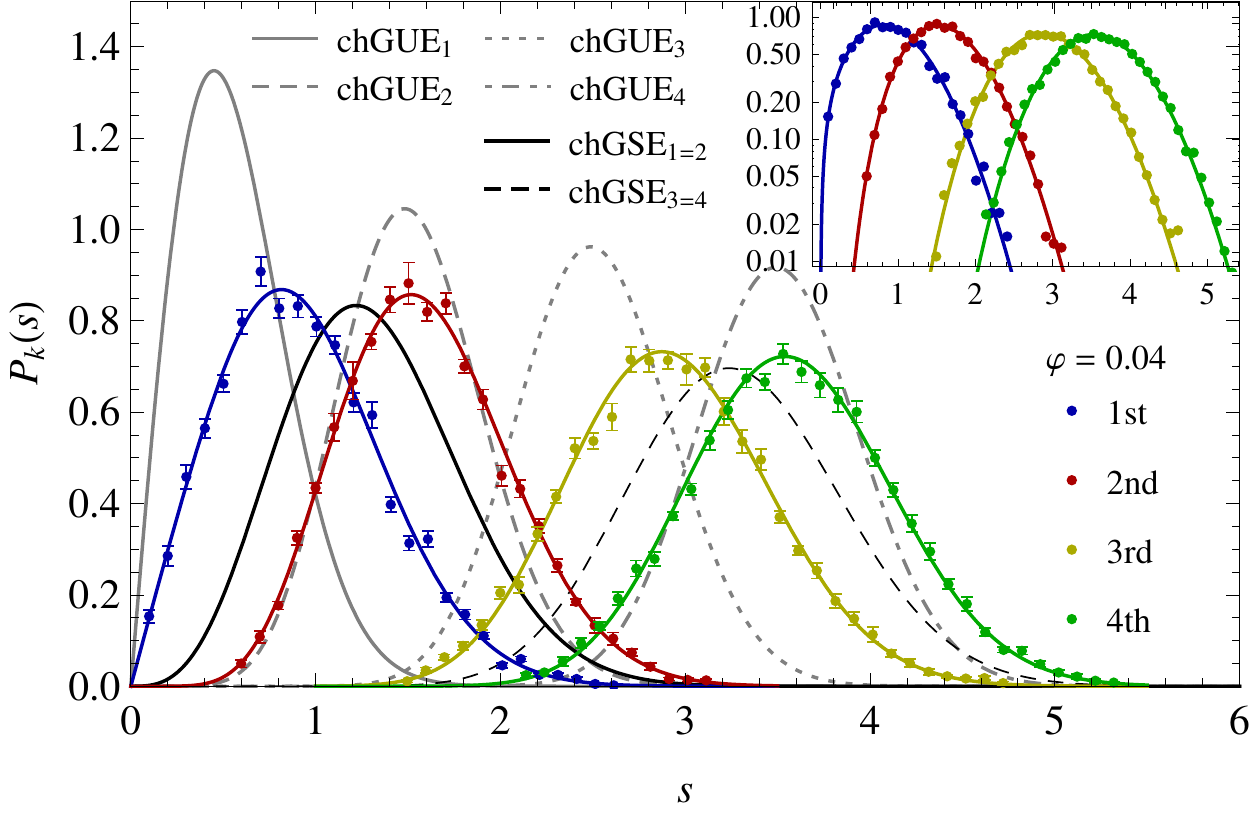}
\caption{ %
\small  
Linear and logarithmic (inset) plots of the first four Dirac eigenvalue distributions $P_{1,2,3,4} (s)$ (blue, red, yellow, green) of
the SU(2)+ICP model on $V=6^4$ and at $\beta = 1.75$,
and best-fit curves of the chGSE--chGUE crossover.
Top left: $\varphi = 0.01$, top right: $\varphi = 0.02$, bottom left: $\varphi = 0.03$, bottom right: $\varphi = 0.04$.
} %
\label{fig:SU2ABF_b175_e010-040_p1-4}
\ \\
\centering 
\includegraphics[width=7.5cm,bb=0 0 360 234]{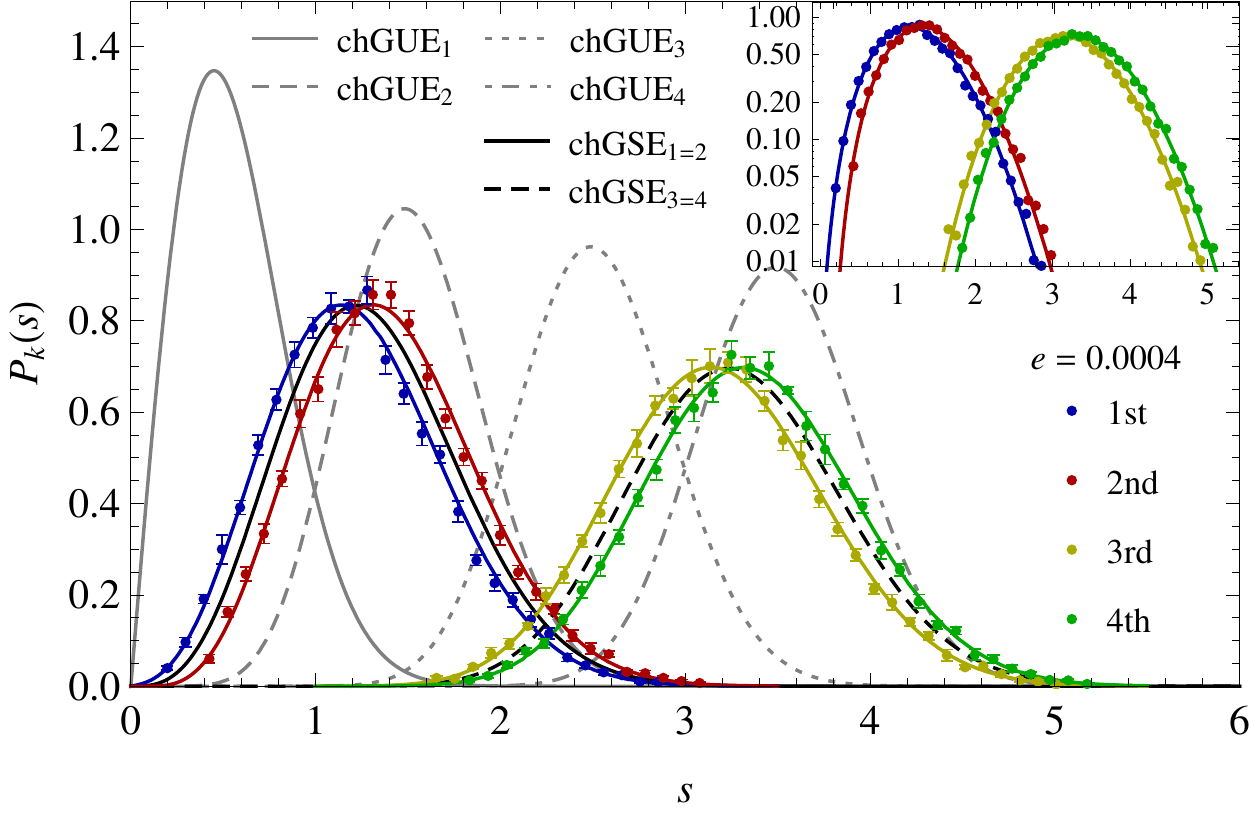}
\includegraphics[width=7.5cm,bb=0 0 360 234]{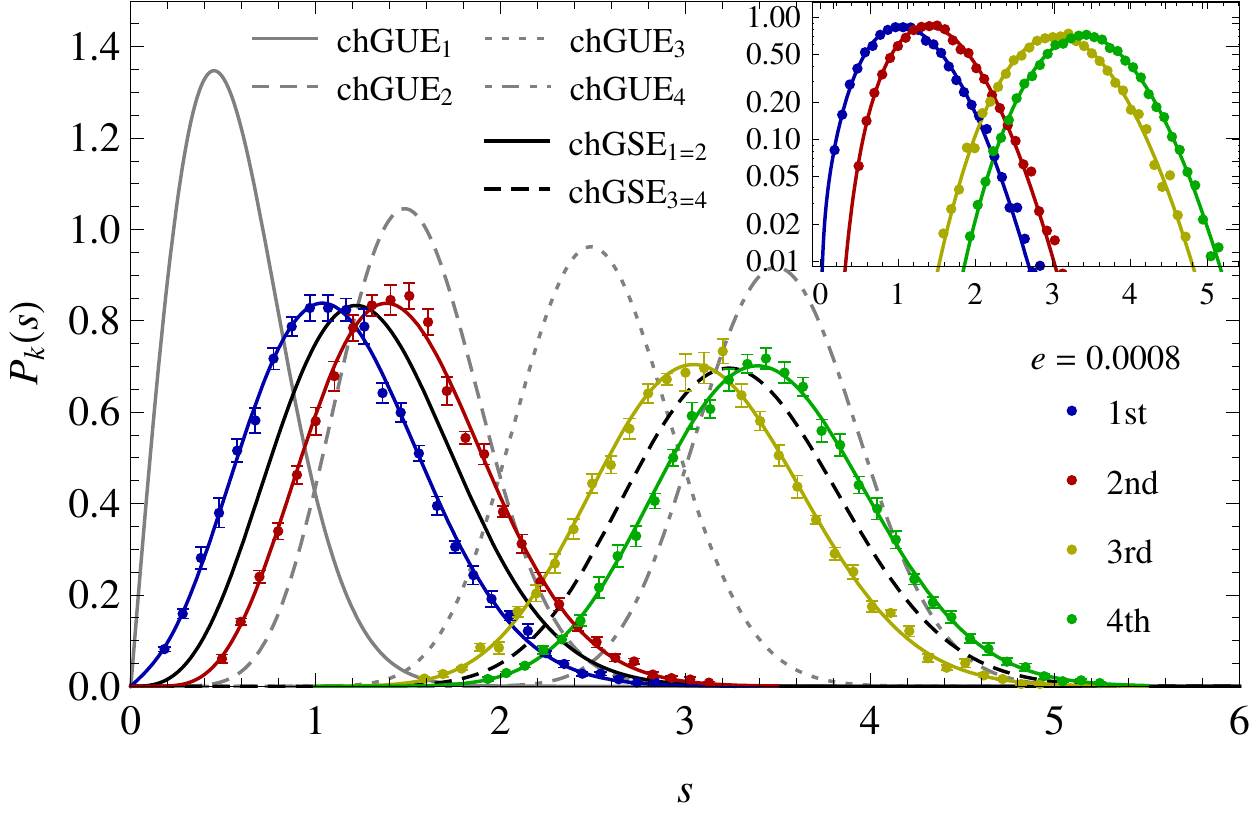}
\includegraphics[width=7.5cm,bb=0 0 360 234]{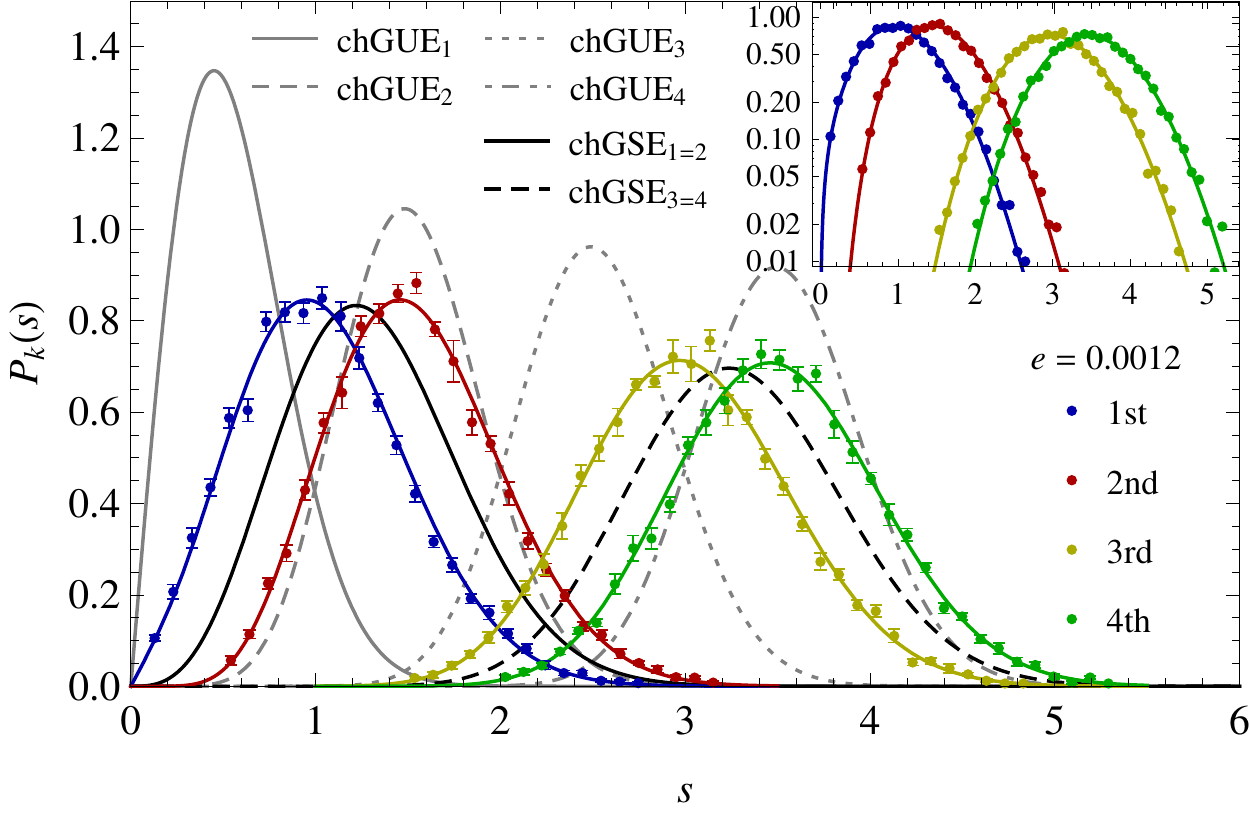}
\includegraphics[width=7.5cm,bb=0 0 360 234]{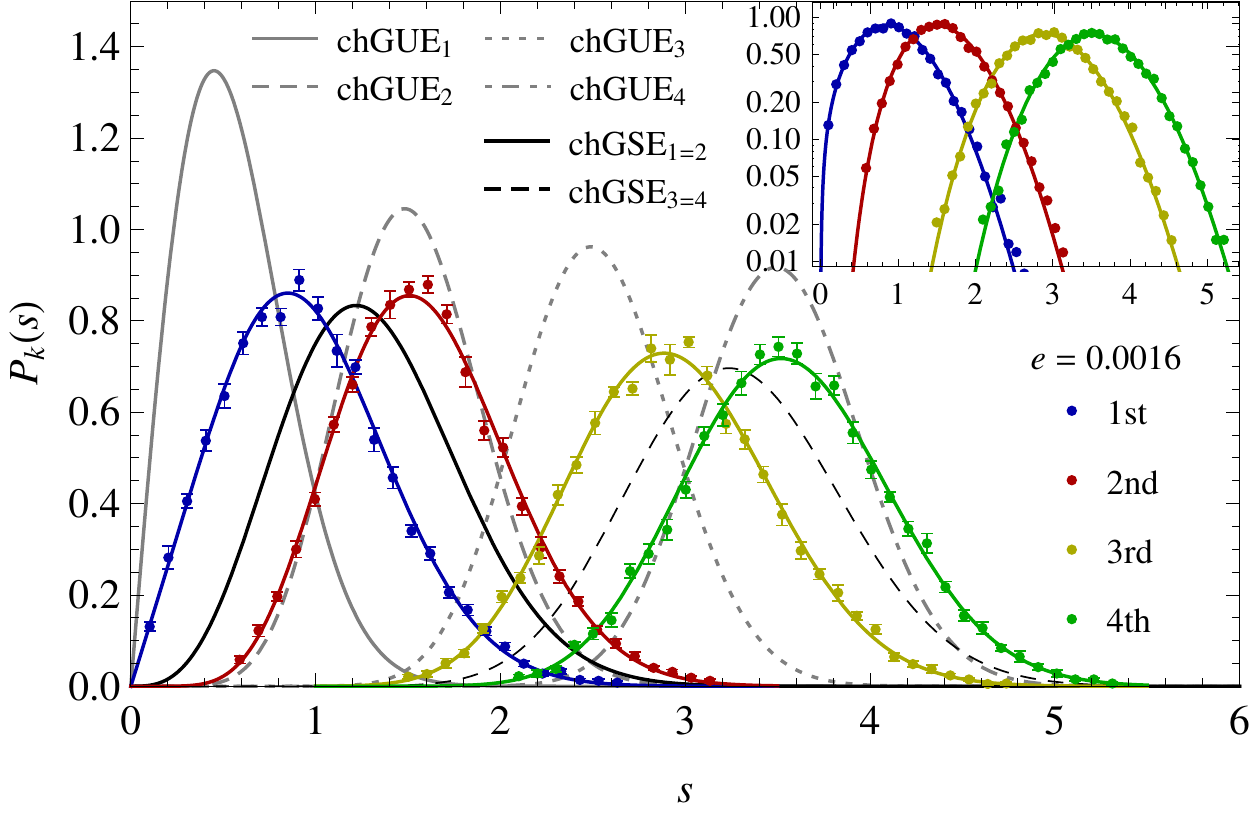}
\caption{ %
\small  
Linear and logarithmic (inset) plots of the first four Dirac eigenvalue distributions $P_{1,2,3,4} (s)$ (blue, red, yellow, green) of
the SU(2)$\times$U(1) model on $V=6^4$ and at $\beta = 1.75$, 
and best-fit curves of the chGSE--chGUE crossover.
Top left: $e= 0.0004$, top right: $e = 0.0008$, 
bottom left: $e = 0.0012$ and bottom right: $e = 0.0016$. 
} %
\label{fig:SU2U1_b175_e0004-0016_p1-4}
\end{figure}

Figure \ref{fig:SU2U1-SU2ABF_b10-20_p1-2} shows samples of the histograms 
$\{P_{1}(s_{1}), P_{2}(s_{2})\}$ of the perturbed Dirac operator eigenvalues and 
best-fit distributions of the chGSE--chGUE crossover.
As the perturbation $\varphi$ or $e$ is increased,
the eigenvalue distributions are clearly seen to follow the random matrix curves.
They respond more rapidly to the perturbation for smaller $\beta$ (left panels) than for 
larger $\beta$ (right panels),
indicating that $F$ is a decreasing function of $\beta$.
For the sake of graphical visibility,
we also exhibit samples of histograms of $\{P_{1}(s_{1}), \ldots , P_{4}(s_{4})\}$ and 
best-fit distributions of the chGSE--chGUE crossover,
at fixed $\beta=1.75$ and with increasing perturbations $\varphi$  (Fig.~\ref{fig:SU2ABF_b175_e010-040_p1-4})
and $e$ (Fig.~\ref{fig:SU2U1_b175_e0004-0016_p1-4}), respectively.

\begin{figure}[H] 
\centering 
\includegraphics[width=6cm,bb=0 0 290 432]{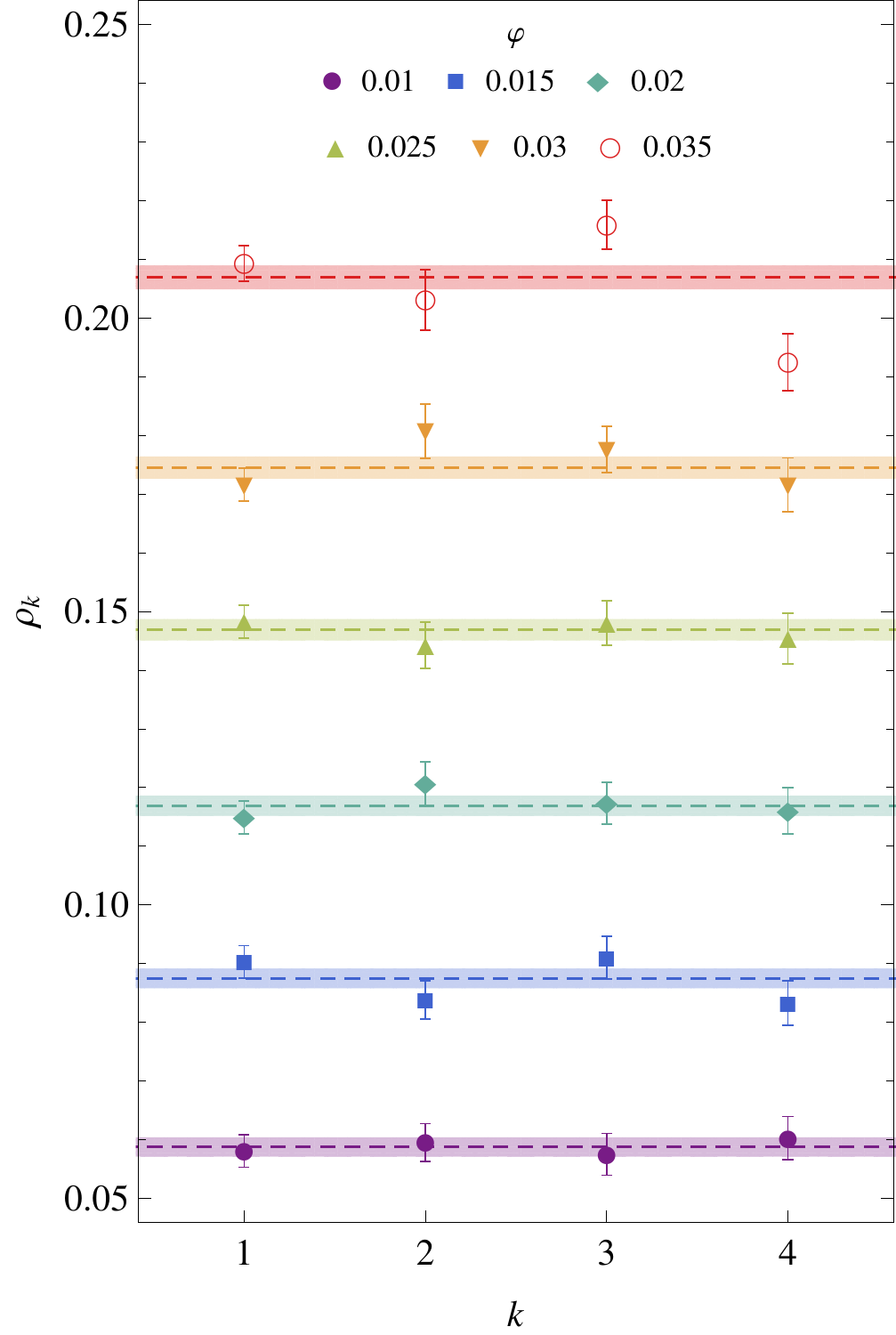}
~~~~
\includegraphics[width=6cm,bb=0 0 290 432]{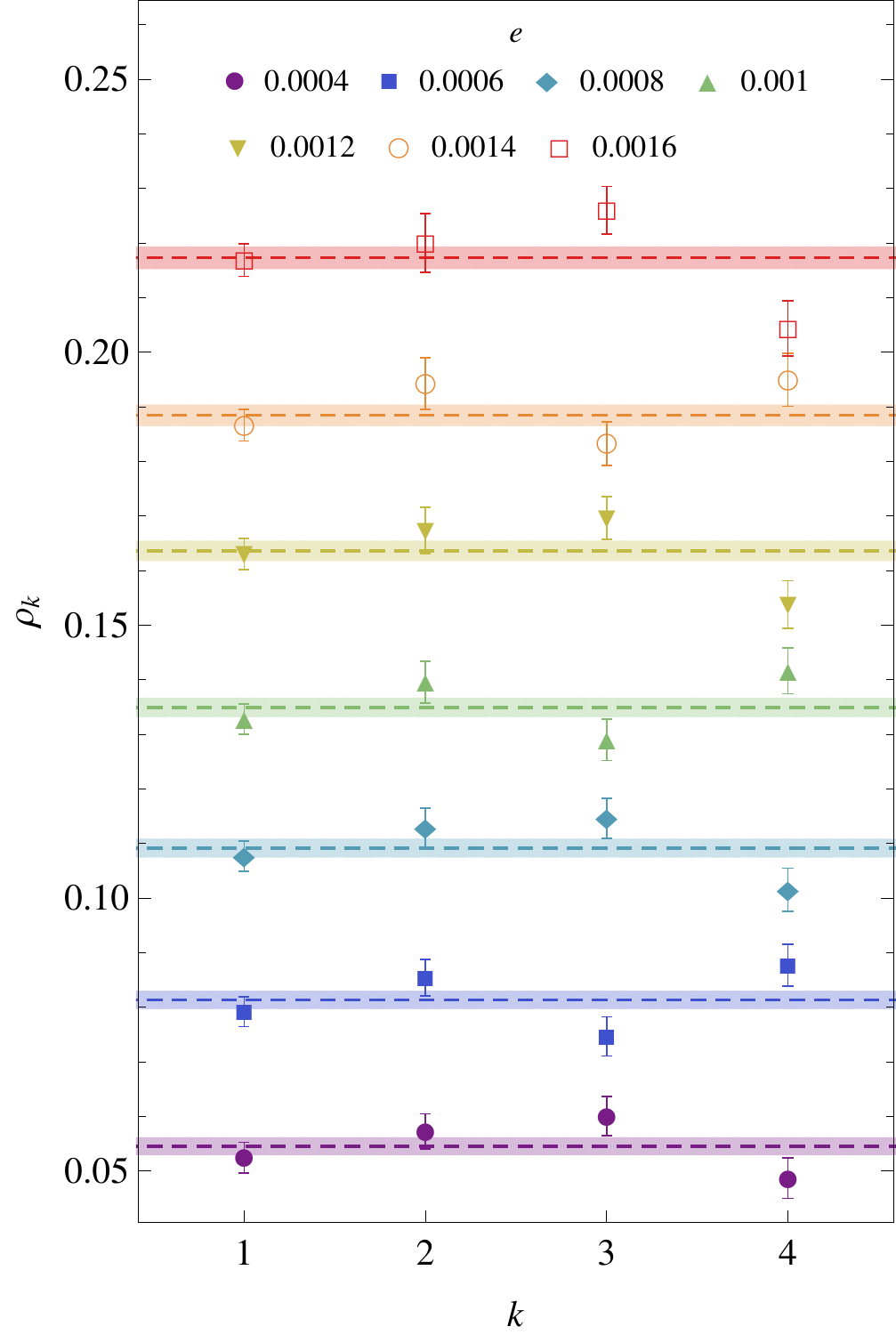}
\caption{ %
Crossover parameters $\rho_{1,2,3,4}$ determined from $P_{1,2,3,4}(s)$.
Their weighted averages and combined errors are shown in horizontal lines and strips.
Left: SU(2)+ICP model with $\varphi = 0.01$--0.035 (purple to red). 
Right: SU(2)$\times$U(1) model with $e= 0.0004$--0.0016 (purple to red). 
Both are at $\beta=1.75$ and on $V=6^4$.
} %
\label{fig:SU2U1-SU2ABF_b175_rhok}
\end{figure}

Figure \ref{fig:SU2U1-SU2ABF_b175_rhok} illustrates
the crossover parameters $\rho_{k}$ determined from $P_k(s_k)$ 
for the SU(2)+ICP model and for the SU(2)$\times$U(1) model at $\beta=1.75$ and $V=6^4$,
at each value of perturbation $\varphi$ or $e$.
We notice that $\rho_{1}$ and $\rho_{2}$ ($\rho_{3}$ and $\rho_{4}$) have a tendency to counter-move
across the weighted average of the four,
as anticipated in Sect.~\ref{simulationdetails}.
Combined use of $P_{2i-1}(s_{2i-1})$ and $P_{2i}(s_{2i})$ is indeed seen to reduce
the errors of the best-fit parameters $\bar{\rho}$ in both panels.

\begin{table}[H] 
\caption{ %
Crossover parameters $\rho$ for the SU(2)+ICP model on $V = 4^4$. 
} %
\label{tab:SU2ABF_n4_rhok} 
\centering 
\begin{tabular}{llllllllc} \hline \hline
~ & ~ & \multicolumn{6}{c}{$\varphi$} & ~  \\
$\beta$ & $\rho$ & 0.01 & 0.02 & 0.03 & 0.04 & 0.05 & 0.06  &  $\chi^2/\text{d.o.f.}$ \\ 
\hline 
0    & $\rho_{1}$ & 0.061(1) & 0.122(1) & 0.183(1) & 0.244(2) & -- & --   & 0.68--1.06     \\   
      & $\rho_{2}$ & 0.061(2) & 0.121(2) & 0.183(2) & 0.238(3) & -- & --  & 0.95--1.71     \\  
      & $\rho_{3}$ & 0.060(2) & 0.118(2) & 0.181(2) & 0.243(2) & -- & --  & 0.50.--0.69     \\  
      & $\rho_{4}$ & 0.062(2) & 0.125(2) & 0.184(2) & 0.245(3) & -- & --  & 0.64--1.19     \\  
  & $\bar{\rho}$ & 0.061\hspace{1.5pt}2(8) & 0.121\hspace{1.5pt}4(9) & 0.183(1) & 0.243(1) & -- & --  & --    \\  \hline  
0.25 & $\rho_{1}$ & 0.060(1) & 0.117(1) & 0.176(1) & 0.234(2) & 0.291(2) & --  & 0.65--1.21     \\    
      & $\rho_{2}$ & 0.057(2) & 0.118(2) & 0.174(2) & 0.234(3) & 0.296(4) & --  & 0.63--1.22     \\     
      & $\rho_{3}$ & 0.062(2) & 0.117(2) & 0.176(2) & 0.235(2) & 0.291(3) & --  & 0.80--1.49    \\    
      & $\rho_{4}$ & 0.056(2) & 0.116(2) & 0.174(2) & 0.231(3) & 0.298(4) & --  & 0.76--1.30     \\    
   & $\bar{\rho}$ & 0.058\hspace{1.5pt}6(8) & 0.116\hspace{1.5pt}9(9) & 0.175\hspace{1.5pt}3(9) & 0.234(1) & 0.292(1) & --   & --    \\    \hline  
0.5  & $\rho_{1}$ & 0.056(1) & 0.108(1) & 0.168(1) & 0.224(2) & 0.279(2) & --  & 0.57--1.56     \\    
      & $\rho_{2}$ & 0.056(2) & 0.117(2) & 0.167(2) & 0.227(3) & 0.282(4) & --  & 0.35--1.33     \\    
      & $\rho_{3}$ & 0.056(2) & 0.111(2) & 0.165(2) & 0.228(2) & 0.279(2) & --  & 0.64--1.25     \\    
      & $\rho_{4}$ & 0.057(2) & 0.112(2) & 0.171(2) & 0.217(3) & 0.277(4) & --  & 0.82--1.34     \\     
   & $\bar{\rho}$ & 0.056\hspace{1.5pt}1(8) & 0.111\hspace{1.5pt}3(9) & 0.167\hspace{1.5pt}7(9) & 0.224(1) & 0.279(1) & --  & --     \\    \hline  
0.75 & $\rho_{1}$ & 0.055(1) & 0.108(1) & 0.162(1) & 0.216(1) & 0.270(2) & --  & 0.63--0.90     \\    
      & $\rho_{2}$ & 0.053(2) & 0.106(2) & 0.159(2) & 0.213(3) & 0.266(4) & --  & 0.76--1.23     \\    
      & $\rho_{3}$ & 0.056(2) & 0.110(2) & 0.164(2) & 0.218(2) & 0.271(2) & --  & 0.78--1.39     \\    
      & $\rho_{4}$ & 0.051(2) & 0.104(2) & 0.158(2) & 0.210(3) & 0.262(3) & --  & 0.98--1.18     \\    
   & $\bar{\rho}$ & 0.053\hspace{1.5pt}8(8) & 0.107\hspace{1.5pt}3(9) & 0.161\hspace{1.5pt}0(9) & 0.215(1) & 0.269(1) & --  & --     \\    \hline  
1.0  & $\rho_{1}$ & 0.052(1) & 0.102(1) & 0.153(1) & 0.203(1) & 0.254(2) & --  & 0.78--1.05     \\    
      & $\rho_{2}$ & 0.050(2) & 0.101(2) & 0.151(2) & 0.201(2) & 0.251(3) & --  & 0.60--1.16     \\    
      & $\rho_{3}$ & 0.050(2) & 0.101(2) & 0.151(2) & 0.202(2) & 0.252(2) & --  & 0.78--1.39     \\    
      & $\rho_{4}$ & 0.051(2) & 0.102(2) & 0.153(2) & 0.204(3) & 0.253(3) & --  & 0.90--1.23     \\    
   & $\bar{\rho}$ & 0.050\hspace{1.5pt}8(8) & 0.101\hspace{1.5pt}5(9) & 0.152\hspace{1.5pt}2(9) & 0.203(1) & 0.253(1) & --  & --     \\    \hline  
1.25 & $\rho_{1}$ & 0.048(1) & 0.096(1) & 0.143(1) & 0.190(1) & 0.238(2) & 0.285(2)  & 0.88--1.23     \\   
      & $\rho_{2}$ & 0.047(2) & 0.095(2) & 0.142(2) & 0.189(2) & 0.236(3) & 0.283(4)  & 0.70--1.14     \\    
      & $\rho_{3}$ & -- & -- & -- & -- & -- & --  & --     \\    
      & $\rho_{4}$ & -- & -- & -- & -- & -- & --  & --     \\     
   & $\bar{\rho}$ & 0.048(1) & 0.095(1) & 0.143(9) & 0.190(1) & 0.237(1) & 0.285(2)  & --     \\    \hline  
1.5  & $\rho_{1}$ & --          & 0.087(1) & 0.130(1) & 0.174(1) & 0.217(2) & 0.261(2)  & 0.69--0.94     \\     
      & $\rho_{2}$ & --          & 0.086(2) & 0.129(2) & 0.172(2) & 0.214(3) & 0.254(3)  & 0.62--1.03     \\    
      & $\rho_{3}$ & -- & -- & -- & -- & -- & --  & --     \\   
      & $\rho_{4}$ & -- & -- & -- & -- & -- & --  & --     \\   
   & $\bar{\rho}$ & --          & 0.087(1) & 0.130(1) & 0.173(1) & 0.216(1) & 0.259(1)  & --     \\    \hline  
1.75$^*$ & $\rho_{1}$ & --          & 0.075(2) & 0.113(2) & 0.150(2) & 0.188(2) & 0.226(2)  & 0.88--1.99     \\    
      & $\rho_{2}$ & --         & 0.074(2) & 0.110(2) & 0.147(3) & 0.180(3) & 0.216(4)  & 0.56--1.30     \\    
      & $\rho_{3}$ & -- & -- & -- & -- & -- & --  &  --    \\   
      & $\rho_{4}$ & -- & -- & -- & -- & -- & --  & --     \\   
   & $\bar{\rho}$ & --         & 0.075(1) & 0.112(2) & 0.149(2) & 0.186(2) & 0.224(2)  & --     \\    
\hline \hline 
\end{tabular}
\end{table}%

\begin{table}[H] 
\caption{ %
Crossover parameters $\rho$ for the SU(2)+ICP model on $V = 6^4$. 
} %
\label{tab:SU2ABF_n6_rhok} 
\centering 
\small{
\begin{tabularx}{153mm}{llXXXXXXXXXc} \hline \hline
~ & ~ & \multicolumn{9}{c}{$\varphi$} & ~  \\
$\beta$ & $\rho$ & 0.01 & 0.015 & 0.02 & 0.025 & 0.03 & 0.035 & 0.04 & 0.045 & 0.05 & $\chi^2/\text{d.o.f.}$ \\ 
\hline 
0    & $\rho_{1}$ & 0.093(3) & 0.135(3) & 0.178(3) & 0.228(3) & -- & --  & -- & -- & --  & 0.54-1.18     \\   
     & $\rho_{2}$ & 0.089(3) & 0.138(4) & 0.187(5) & 0.228(6) & -- & -- & -- & -- & --  & 0.47--1.40     \\  
     & $\rho_{3}$ & 0.083(4) & 0.135(4) & 0.173(4) & 0.223(4) & -- & -- & -- & -- & --  & 0.56--1.30     \\  
     & $\rho_{4}$ & 0.099(4) & 0.138(4) & 0.191(5) & 0.236(6) & -- & -- & -- & -- & --  & 0.76--1.09     \\  
  & $\bar{\rho}$ & 0.091(2) & 0.136(2) & 0.180(2) & 0.228(2) & -- & -- & -- & -- & --  & --     \\   \hline 
0.25 & $\rho_{1}$ & 0.088(3) & 0.135(3) & 0.177(3) & 0.223(3) & -- & -- & -- & -- & --  & 0.87--1.32     \\   
      & $\rho_{2}$ & 0.087(3) & 0.130(4) & 0.171(4) & 0.215(5) & -- & -- & -- & -- & --  & 0.51--1.37     \\  
      & $\rho_{3}$ & 0.090(4) & 0.133(4) & 0.179(4) & 0.218(4) & -- & -- & -- & -- & --  & 0.66--1.05     \\  
      & $\rho_{4}$ & 0.087(4) & 0.132(4) & 0.169(4) & 0.229(6) & -- &  -- & -- & -- & --  & 0.76--1.19     \\  
   & $\bar{\rho}$ & 0.088(2) & 0.133(2) & 0.175(2) & 0.221(2) & -- & -- & -- & -- & --  & --     \\   \hline 
0.5  & $\rho_{1}$ & 0.083(3) & 0.119(3) & 0.168(3) & 0.217(3) & -- & -- & -- & -- & --  & 0.96--1.25     \\   
      & $\rho_{2}$ & 0.085(3) & 0.135(4) & 0.170(4) & 0.201(5) & -- & -- & -- & -- & --  & 0.60--0.94     \\  
      & $\rho_{3}$ & 0.083(4) & 0.123(4) & 0.166(4) & 0.222(4) & -- & -- & -- & -- & --  & 0.77--1.32     \\  
      & $\rho_{4}$ & 0.086(4) & 0.130(4) & 0.173(5) & 0.200(5) & -- & -- & -- & -- & --  & 0.48--1.23     \\  
   & $\bar{\rho}$ & 0.084(2) & 0.125(2) & 0.169(2) & 0.213(2) & -- & -- & -- & -- & -- & --     \\   \hline 
0.75 & $\rho_{1}$ & 0.082(3) & 0.123(3) & 0.157(3) & 0.202(3) & 0.240(3) & -- & -- & -- & --  & 0.70--1.47     \\   
      & $\rho_{2}$ & 0.077(3) & 0.119(4) & 0.171(4) & 0.205(5) & 0.246(6) & -- & -- & -- & --  & 0.46--1.12     \\  
      & $\rho_{3}$ & 0.087(4) & 0.124(4) & 0.163(4) & 0.207(4) & 0.241(4) & -- & -- & -- & --  & 0.60--1.14     \\    
      & $\rho_{4}$ & 0.076(4) & 0.116(4) & 0.159(4) & 0.200(5) & 0.244(6) & -- & -- & -- & --  & 0.80--1.33     \\    
  & $\bar{\rho}$ & 0.081(2) & 0.121(2) & 0.161(2) & 0.203(2) & 0.242(2) & -- & -- & -- & --  & --     \\     \hline 
1.0  & $\rho_{1}$ & 0.076(3) & 0.111(3) & 0.153(3) & 0.188(3) & 0.233(3) & -- & -- & -- & --  & 0.49--0.97     \\   
      & $\rho_{2}$ & 0.080(3) & 0.117(4) & 0.157(4) & 0.197(5) & 0.228(6) & -- & -- & -- & --  & 0.71--1.27     \\   
      & $\rho_{3}$ & 0.078(4) & 0.108(4) & 0.146(4) & 0.186(4) & 0.227(4) & -- & -- & -- & --  & 0.75--1.38     \\    
      & $\rho_{4}$ & 0.078(4) & 0.125(4) & 0.165(5) & 0.199(5) & 0.233(6) & -- & -- & -- & --  & 0.70--1.13     \\    
   & $\bar{\rho}$ & 0.078(2) & 0.114(2) & 0.154(2) & 0.191(2) & 0.231(2) & -- & -- & -- & --  & --     \\     \hline 
1.25 &$\rho_{1}$ & 0.074(3) & 0.107(3) & 0.146(3) & 0.180(3) & 0.213(3) & -- & -- & -- & --  & 0.54--1.31     \\    
     & $\rho_{2}$ & 0.073(3) & 0.113(4) & 0.144(4) & 0.180(5) & 0.219(5) & -- & -- & -- & --  & 0.67--1.05     \\    
     & $\rho_{3}$ & 0.071(4) & 0.106(4) & 0.152(4) & 0.181(4) & 0.214(4) & -- & -- & -- & --  & 0.84--0.98     \\    
     & $\rho_{4}$ & 0.074(4) & 0.111(4) & 0.138(4) & 0.179(5) & 0.222(5) & -- & -- & -- & --  & 0.66--1.24     \\    
  & $\bar{\rho}$ & 0.073(2) & 0.109(2) & 0.145(2) & 0.180(2) & 0.216(2) & -- & -- & -- & --  & --     \\     \hline 
1.5  & $\rho_{1}$ & 0.068(3) & 0.099(3) & 0.136(3) & 0.169(3) & 0.196(3) & 0.230(3) & -- & -- & --  & 0.6--0.89     \\    
      & $\rho_{2}$ & 0.067(3) & 0.104(4) & 0.132(4) & 0.158(4) & 0.206(5) & 0.240(6) & -- & -- & --  & 0.60--1.02     \\    
      & $\rho_{3}$ & 0.069(4) & 0.106(4) & 0.141(4) & 0.172(4) & 0.201(4) & 0.232(4) & -- & -- & --  & 0.80--1.17     \\    
      & $\rho_{4}$ & 0.063(4) & 0.094(4) & 0.128(4) & 0.157(4) & 0.200(5) & 0.231(6) & -- & -- & --  & 0.84--1.32     \\    
   & $\bar{\rho}$ & 0.067(2) & 0.100(2) & 0.135(2) & 0.166(2) & 0.199(2) & 0.232(2) & -- & -- & --  & --     \\     \hline 
1.75 & $\rho_{1}$ & 0.058(3) & 0.090(3) & 0.115(3) & 0.148(3) & 0.172(3) & 0.209(3) & -- & -- & --  & 0.67--1.21     \\    
      & $\rho_{2}$ & 0.060(3) & 0.084(3) & 0.121(4) & 0.144(4) & 0.181(5) & 0.203(5) & -- & -- & --  & 0.67--1.60     \\    
      & $\rho_{3}$ & 0.057(4) & 0.091(4) & 0.117(4) & 0.148(4) & 0.178(4) & 0.216(4) & -- & -- & --  & 0.76--1.54     \\    
      & $\rho_{4}$ & 0.060(4) & 0.083(4) & 0.116(4) & 0.145(4) & 0.172(5) & 0.192(5) & -- & -- & --  & 0.77--1.68     \\    
   & $\bar{\rho}$ & 0.059(2) & 0.087(2) & 0.117(2) & 0.147(2) & 0.175(2) & 0.207(2) & -- & -- & --  & --     \\     \hline 
2.0   & $\rho_{1}$ & --          & 0.072(3) & 0.095(3) & 0.117(3) & 0.140(3) & 0.163(3) & 0.185(3) & 0.207(3) & 0.230(3)  & 0.94--1.48     \\    
       & $\rho_{2}$ & --          & 0.063(3) & 0.085(3) & 0.108(4) & 0.131(4) & 0.152(4) & 0.173(4) & 0.194(5) & 0.214(5)  & 0.90--1.37     \\    
       & $\rho_{3}$ & -- & -- & -- & -- & -- & -- & -- & -- & --  & --     \\    
       & $\rho_{4}$ & -- & -- & -- & -- & -- & -- & -- & -- & --  & --     \\    
   & $\bar{\rho}$ & -- & 0.068(2) & 0.091(2) & 0.114(2) & 0.137(2) & 0.159(2) & 0.181(2) & 0.204(3) & 0.226(3)  & --     \\   \hline 
2.1$^*$   & $\rho_{1}$ & -- & -- & 0.079(4) & -- & 0.114(4) & 0.133(4) & 0.151(4) & 0.170(4) & 0.189(4)  & 1.55--1.90     \\   
       & $\rho_{2}$ & -- & -- & 0.064(5) & - & 0.101(5) & 0.119(5) & 0.137(5) & 0.150(6) & 0.167(6)  & 1.42--1.94     \\    
       & $\rho_{3}$ & -- & -- & -- & -- & -- & -- & -- & -- & --  & --     \\    
       & $\rho_{4}$ & -- & -- & -- & -- & -- & -- & -- & -- & --  & --     \\    
   & $\bar{\rho}$ & -- & -- & 0.073(3) & -- & 0.109(3) & 0.128(3) & 0.147(3) & 0.165(3) & 0.183(3)  & --    \\  
\hline \hline 
\end{tabularx}}
\end{table}%

\begin{table}[H] 
\caption{%
Crossover parameters $\rho$ for the SU(2)$\times$U(1) model on $V = 4^4$. 
}%
\label{tab:SU2U1_n4_rhok}  
\centering 
\small{
\begin{tabularx}{153mm}{lllllllllc} \hline \hline
~ & ~ & \multicolumn{7}{c}{$e$} & ~   \\
$\beta$ & $\rho$ & 0.002 & 0.003 & 0.004 & 0.005 & 0.006 & 0.008 & 0.010  & $\chi^2/\text{d.o.f.}$  \\ 
\hline 
0    & $\rho_{1}$ & 0.093(1) & 0.136(1) & 0.186(1) & 0.232(2) & 0.276(2) & -- & --   & 0.62--1.52     \\   
      & $\rho_{2}$ & 0.905(2) & 0.142(2) & 0.181(2) & 0.225(3) & 0.277(4) & -- & --    & 0.73--1.38     \\   
      & $\rho_{3}$ & 0.089(2) & 0.136(2) & 0.183(2) & 0.231(2) & 0.273(2) & -- & --   & 0.82--1.20     \\   
      & $\rho_{4}$ & 0.094(2) & 0.139(2) & 0.187(2) & 0.230(3) & 0.280(4) & -- & --   & 0.81--1.08     \\   
  & $\bar{\rho}$ & 0.091\hspace{1.5pt}8(8) & 0.137\hspace{1.5pt}8(9) & 0.184(1) & 0.230(1) & 0.276(1) & -- & --    & --     \\     \hline  
0.25 & $\rho_{1}$ & 0.088(1) & 0.136(1) & 0.176(1) & 0.225(2) & 0.265(2) & -- & --   & 0.48--1.14     \\   
      & $\rho_{2}$ & 0.088(2) & 0.128(2) & 0.177(2) & 0.216(3) & 0.264(3) & -- & --   & 0.76--1.75     \\   
      & $\rho_{3}$ & 0.087(2) & 0.137(2) & 0.175(2) & 0.226(2) & 0.262(2) & -- & --   & 0.55--1.04     \\   
      & $\rho_{4}$ & 0.088(2) & 0.128(2) & 0.176(2) & 0.214(3) & 0.261(3) & -- & --   & 0.75--1.53     \\   
   & $\bar{\rho}$ & 0.088\hspace{1.5pt}0(8) & 0.133\hspace{1.5pt}1(9) & 0.176\hspace{1.5pt}0(9) & 0.222(1) & 0.264(1) & -- & --    & --     \\    \hline  
0.5  & $\rho_{1}$ & 0.083(1) & 0.126(1) & 0.167(1) & 0.210(2) & 0.251(2) & -- & -- & 0.50--0.99     \\   
      & $\rho_{2}$ & 0.086(2) & 0.128(2) & 0.170(2) & 0.212(3) & 0.256(3) & -- & --     & 0.53--0.94     \\   
      & $\rho_{3}$ & 0.086(2) & 0.121(2) & 0.170(2) & 0.206(2) & 0.255(2) & -- & --   & 0.79--1.34     \\    
      & $\rho_{4}$ & 0.082(2) & 0.133(2) & 0.166(2) & 0.218(3) & 0.248(3) & -- & --   & 0.38--1.34     \\    
   & $\bar{\rho}$ & 0.084\hspace{1.5pt}0(8) & 0.126\hspace{1.5pt}2(9) & 0.168\hspace{1.5pt}2(9) & 0.210(1) & 0.252(1) & -- & --    & --     \\    \hline  
0.75 & $\rho_{1}$ & 0.078(1) & 0.118(1) & 0.159(1) & 0.199(1) & 0.239(2) & -- & --   & 0.68--1.22     \\   
      & $\rho_{2}$ & 0.082(2) & 0.122(2) & 0.163(2) & 0.202(2) & 0.244(3) & -- & --     & 0.48--0.84     \\   
      & $\rho_{3}$ & 0.081(2) & 0.116(2) & 0.161(2) & 0.196(2) & 0.242(2) & -- & --    & 0.72--1.01     \\   
      & $\rho_{4}$ & 0.080(2) & 0.124(2) & 0.159(2) & 0.204(3) & 0.237(3) & -- & --  & 0.62--1.40     \\    
   & $\bar{\rho}$ & 0.079\hspace{1.5pt}9(8) & 0.119\hspace{1.5pt}8(9) & 0.160\hspace{1.5pt}2(9) & 0.200(1) & 0.240(1) & -- & --   & --     \\    \hline  
1.0  & $\rho_{1}$ & 0.075(1) & 0.114(1)  & 0.150(1) & 0.189(1)  & 0.225(2) & 0.300(2) & --   & 0.75--1.27      \\    
      & $\rho_{2}$ & 0.075(2) & 0.111(2)  & 0.150(2) & 0.184(2)  & 0.223(3) & 0.296(4) & --    & 1.00--1.55     \\     
      & $\rho_{3}$ & 0.076(2) & 0.114(2)  & 0.151(2) & 0.189(2)  & 0.226(2) & 0.303(3) & --    & 0.94--1.62     \\     
      & $\rho_{4}$ & 0.073(2) & 0.111(2)  & 0.147(2) & 0.183(2)  & 0.218(2) & 0.283(4) & --  & 0.86--1.20     \\     
   & $\bar{\rho}$ & 0.074\hspace{1.5pt}8(8) & 0.112\hspace{1.5pt}6(9) & 0.149\hspace{1.5pt}4(9) & 0.187(1) & 0.224(1) & 0.298(1) & --  & --        \\    \hline  
1.25 & $\rho_{1}$ & 0.068(1) & 0.103(1)  & 0.138(1) & 0.172(1)  & 0.208(1) & 0.278(2) & --     & 0.60--1.15     \\   
      & $\rho_{2}$ & 0.069(2) & 0.104(2)  & 0.138(2) & 0.172(2) & 0.206(3) & 0.272(4) & --  & 0.53--0.95     \\   
      & $\rho_{3}$ & 0.069(2) & 0.107(2)  & 0.139(2) & 0.177(2)  & 0.210(2) & 0.281(2) & --   & 1.14--1.57     \\   
      & $\rho_{4}$ & 0.069(2) & 0.097(2)  & 0.134(2) & 0.162(2)  & 0.201(3) & 0.264(3) & --    & 1.32--1.65     \\     
   & $\bar{\rho}$ & 0.068\hspace{1.5pt}8(8) & 0.103\hspace{1.5pt}2(8) & 0.137\hspace{1.5pt}7(9) & 0.171\hspace{1.5pt}6(9) & 0.207(1) & 0.276(1) & --   & --     \\     \hline  
1.5  & $\rho_{1}$ & 0.063(1) & 0.092(1)  & 0.125(1) & 0.154(1)  & 0.187(1) & 0.250(2) & 0.313(2)   & 0.68--1.06     \\     
      & $\rho_{2}$ & 0.060(2) & 0.093(2)  & 0.121(2) & 0.154(2)  & 0.180(2) & 0.236(3) & 0.286(4)    & 0.61--1.25     \\    
      & $\rho_{3}$ & -- & -- & -- & -- & -- & -- & --   & --     \\   
      & $\rho_{4}$ & -- & -- & -- & -- & -- & -- & --   & --     \\   
   & $\bar{\rho}$ & 0.062(1) & 0.092(1)  & 0.123(1) & 0.154(1)  & 0.185(1) & 0.246(1) & 0.308(2)  & --     \\    \hline  
1.75$^*$ & $\rho_{1}$ & 0.054(2) & --          & 0.106(2) & --          & 0.158(2) & 0.210(2) & 0.263(2)  & 1.35--1.63     \\    
      & $\rho_{2}$ & 0.049(2) & --          & 0.100(2) & --          & 0.148(2) & 0.195(3) & 0.236(3)   & 1.16--1.77     \\    
      & $\rho_{3}$ & -- & -- & -- & -- & -- & -- & --  & --     \\   
      & $\rho_{4}$ & -- & -- & -- & -- & -- & -- & --   & --     \\   
   & $\bar{\rho}$ & 0.052(1) & --          & 0.103(1) & --          & 0.154(1) & 0.206(1) & 0.258(2)   & --    \\   
\hline \hline 
\end{tabularx}}
\end{table}%

\begin{table}[H] 
\caption{%
Crossover parameters $\rho$ for the SU(2)$\times$U(1) model on $V = 6^4$. 
}%
\label{tab:SU2U1_n6_rhok} 
\centering 
\footnotesize{
\begin{tabularx}{153mm}{llXXXXXXXXXXc} \hline \hline
~ & ~ & \multicolumn{9}{c}{$e$} & ~ \\
$\beta$ & $\rho$ & 0.0004 & 0.0006 & 0.0008 & 0.0010 & 0.0012 & 0.0014 & 0.0016 & 0.0020 & 0.0024 & 0.0028 & $\chi^2/\text{d.o.f.}$ \\ 
\hline 
0    & $\rho_{1}$ & 0.090(3) & 0.149(3) & 0.192(3) & 0.241(3) & -- & -- & --  & -- & -- & -- & 0.72--1.03 \\ 
     & $\rho_{2}$ & 0.102(4) & 0.134(4) & 0.187(5) & 0.232(6) & -- & --  & -- & -- & -- & --  & 0.97--1.31     \\  
     & $\rho_{3}$ & 0.098(4) & 0.147(4) & 0.196(4) & 0.236(4) & -- & -- & -- & -- & -- & --  & 0.67--1.45     \\  
     & $\rho_{4}$ & 0.092(4) & 0.138(4) & 0.183(5) & 0.240(6) & -- & -- & -- & -- & -- & --  & 0.59--1.09     \\  
  & $\bar{\rho}$ & 0.095(2) & 0.143(2) & 0.190(2) & 0.238(2) & -- & -- & -- & -- & -- & --  & --    \\  \hline 
0.25 & $\rho_{1}$ & 0.093(3) & 0.134(3) & 0.185(3) & 0.225(3) & -- & -- & -- & -- & -- & --  & 0.49--1.21     \\   
      & $\rho_{2}$ & 0.089(3) & 0.143(4) & 0.180(5) & 0.238(6) & -- & -- & -- & -- & -- & --  & 0.97--1.41     \\  
      & $\rho_{3}$ & 0.093(4) & 0.132(4) & 0.184(4) & 0.223(4) & -- & -- & -- & -- & -- & --  & 0.67--1.31     \\  
      & $\rho_{4}$ & 0.090(4) & 0.147(4) & 0.180(5) & 0.243(6) & -- & -- & -- & -- & -- & --  & 1.01--1.51     \\  
   & $\bar{\rho}$ & 0.092(2) & 0.138(2) & 0.183(2) & 0.228(2) & -- & -- & -- & -- & -- & --  & --     \\  \hline 
0.5  & $\rho_{1}$ & 0.090(3) & 0.127(3) & 0.177(3) & 0.215(3) & -- & --  & -- & -- & -- & --  & 0.73--1.19     \\   
      & $\rho_{2}$ & 0.085(3) & 0.140(4) & 0.173(4) & 0.230(6) & -- & -- & -- & -- & -- & --  & 0.57--1.20     \\  
      & $\rho_{3}$ & 0.085(4) & 0.128(4) & 0.173(4) & 0.214(4) & -- & --  & -- & -- & -- & --  & 0.64--1.10     \\  
      & $\rho_{4}$ & 0.091(4) & 0.135(4) & 0.180(5) & 0.222(5) & -- & -- & -- & -- & -- & --  & 0.82--0.98     \\  
   & $\bar{\rho}$ & 0.088(2) & 0.131(2) & 0.176(2) & 0.218(2) & -- & -- & -- & -- & -- & -- & --    \\  \hline 
0.75 & $\rho_{1}$ & 0.082(3) & 0.124(3) & 0.166(3) & 0.206(3)  & 0.249(3) & -- & -- & -- & -- & --  & 0.77--0.96     \\   
      & $\rho_{2}$ & 0.084(3) & 0.128(4) & 0.167(4) & 0.213(5)  & 0.250(6) & -- & -- & -- & -- & --  & 0.74--0.79     \\  
      & $\rho_{3}$ & 0.082(4) & 0.123(4) & 0.166(4) & 0.205(4)  & 0.248(5) & -- & -- & -- & -- & --  & 0.84--1.02     \\    
      & $\rho_{4}$ & 0.082(4) & 0.129(4) & 0.163(4) & 0.214(5)  & 0.244(6) & -- & -- & -- & -- & --  & 0.94--1.12     \\    
  & $\bar{\rho}$ & 0.082(2) & 0.125(2) & 0.166(2) & 0.208(2)  & 0.248(2) & -- & -- & -- & -- & --  & --     \\    \hline 
1.0  & $\rho_{1}$ & 0.081(3) & 0.116(3) & 0.155(3) & 0.193(3)  & 0.232(3) & -- & -- & -- & -- & --  & 1.02--1.36     \\   
      & $\rho_{2}$ & 0.073(3) & 0.117(4) & 0.154(4) & 0.194(5)  & 0.231(6) & -- & -- & -- & -- & --  & 0.75--1.01     \\   
      & $\rho_{3}$ & 0.083(4) & 0.113(4) & 0.161(4) & 0.191(4)  & 0.241(5) & -- & -- & -- & -- & --  & 0.36--1.45     \\    
      & $\rho_{4}$ & 0.072(4) & 0.119(4) & 0.155(4) & 0.196(5)  & 0.234(6) & -- & -- & -- & -- & --  & 0.75--1.09     \\    
   & $\bar{\rho}$ & 0.078(2) & 0.116(2) & 0.156(2) & 0.193(2)  & 0.234(2) & -- & -- & -- & -- & --  & --     \\    \hline 
1.25 &$\rho_{1}$ & 0.070(3) & 0.108(3)  & 0.143(3) & 0.178(3)  & 0.216(3) & -- & -- & -- & -- & --  & 0.70--1.13     \\    
     & $\rho_{2}$ & 0.076(3) & 0.110(4)  & 0.149(4) & 0.182(5)  & 0.222(6) & -- & -- & -- & -- & --  & 0.76--0.95     \\    
     & $\rho_{3}$ & 0.073(4) & 0.111(4)  & 0.144(4) & 0.184(4)  & 0.217(4) & -- & -- & -- & -- & --  & 0.87--1.23     \\    
     & $\rho_{4}$ & 0.071(4) & 0.108(4)  & 0.143(4) & 0.179(5)  & 0.213(5) & -- & -- & -- & -- & --  &  0.48--0.94    \\    
  & $\bar{\rho}$ & 0.072(2) & 0.109(2) & 0.144(2) & 0.180(2)  & 0.217(2) & -- & -- & -- & -- & --  & --     \\    \hline 
1.5  & $\rho_{1}$ & 0.062(3) & 0.095(3) & 0.130(3) & 0.160(3)  & 0.195(3) & -- & -- & -- & -- & --  & 0.77--1.33     \\    
      & $\rho_{2}$ & 0.068(3) & 0.098(3)  & 0.132(4) & 0.165(4)  & 0.196(5) & -- & -- & -- & -- & --  & 0.68--1.26     \\    
      & $\rho_{3}$ & 0.069(4) & 0.100(4)  & 0.129(4) & 0.164(4)  & 0.193(4) & -- & -- & -- & -- & --  & 0.78--1.25     \\    
      & $\rho_{4}$ & 0.059(4) & 0.094(4) & 0.127(4) & 0.157(4)  & 0.189(5) & -- & -- & -- & -- & --  & 0.96--0.99     \\    
   & $\bar{\rho}$ & 0.064(2) & 0.097(2)  & 0.130(2) & 0.161(2)  & 0.194(2) & -- & -- & -- & -- & --  & --     \\    \hline 
1.75 & $\rho_{1}$ & 0.052(3) & 0.079(3)  & 0.108(3) & 0.133(3)  & 0.163(3) & 0.187(3) & 0.217(3) & -- & -- & --  & 0.93--1.24     \\    
      & $\rho_{2}$ & 0.057(3) & 0.085(3)  & 0.113(4) & 0.140(4)  & 0.167(4) & 0.194(5) & 0.220(5) & -- & -- & --  & 0.87--1.20     \\    
      & $\rho_{3}$ & 0.060(4) & 0.075(4)  & 0.115(4) & 0.129(4)  & 0.170(4) & 0.183(4) & 0.226(4) & -- & -- & --  & 0.66--1.22     \\    
      & $\rho_{4}$ & 0.049(4) & 0.088(4)  & 0.102(4) & 0.141(4)  & 0.154(4) & 0.195(5) & 0.204(5) & -- & -- & --  & 0.46--1.09     \\    
   & $\bar{\rho}$ & 0.055(2) & 0.081(2)  & 0.109(2) & 0.135(2)  & 0.163(2) & 0.188(2) & 0.217(2) & -- & -- & --  & --     \\    \hline 
2.0   & $\rho_{1}$ & --         & 0.057(3)  & 0.080(3) & 0.096(3)  & 0.118(3) & 0.134(3) & 0.156(3) & 0.194(3) & 0.234(3) & 0.272(3)  & 0.95--1.44     \\    
       & $\rho_{2}$ & --         & 0.058(3)  & 0.073(3) & 0.095(3)  & 0.110(4) & 0.132(4) & 0.147(4) & 0.180(5) & 0.215(5) & 0.244(6) & 0.90--1.94     \\    
       & $\rho_{3}$ & -- & -- & -- & -- & -- & -- & -- & -- & --  & -- & --     \\    
       & $\rho_{4}$ & -- & -- & -- & -- & -- & -- & -- & -- & --  & -- & --     \\    
   & $\bar{\rho}$ & -- & 0.57(2) & 0.077(2) & 0.096(2) & 0.115(2) & 0.133(2) & 0.153(2) & 0.190(2) & 0.229(3) & 0.266(3) & --      \\   \hline 
2.1$^*$   & $\rho_{1}$ & -- & -- & 0.062(4) & -- & 0.091(4) & -- & -- & 0.147(4) & 0.176(4) & 0.205(4)  & 1.03--1.56     \\   
       & $\rho_{2}$ & -- & -- & 0.059(5) & -- & 0.087(5) & -- & -- & 0.139(5) & 0.162(6) & 0.187(7)  & 1.29--1.76     \\    
       & $\rho_{3}$ & -- & -- & -- & -- & -- & -- & -- & -- & -- & --  & --     \\    
       & $\rho_{4}$ & -- & -- & -- & -- & -- & -- & -- & -- & -- & --  & --     \\    
   & $\bar{\rho}$ & -- & -- & 0.061(3) & -- & 0.089(3) & -- & -- & 0.144(3) & 0.172(3) & 0.201(3)  & --    \\  
\hline \hline 
\end{tabularx}}%
\end{table}%

\subsubsection{Low-energy constants} 
The chiral condensate $\Sigma$ is determined by Eq.~\eqref{chiralcondensate}
from $\bar{\varDelta}$ summarized in Tables \ref{tab:SU2ABF_mls} and \ref{tab:SU2U1_mls}.
In Table \ref{tab:low-energy-constants} 
we list these values
in the third column (SU(2)+ICP) and 
in the fifth column (SU(2)$\times$U(1)),
at each $\beta$ and on $V=4^4$ and $6^4$ lattices.
In addition, 
the values of the chiral condensate in the thermodynamic limit $V=\infty$, 
extrapolated from $V=4^4$ and $6^4$, are listed for the coupling range $0 \leq \beta \leq 1.75$.\footnote{
Although the fitting range of $P_k(s)$ at $\beta=1.75$ on $V=4^4$ is compromised as compared to $\beta\leq 1.5$,
we dare to estimate the thermodynamic limit of the chiral condensate and pseudo-scalar decay constant
using these data.}

\begin{figure}[H] 
\centering 
\includegraphics[width=6.0cm,bb=0 0 270 432]{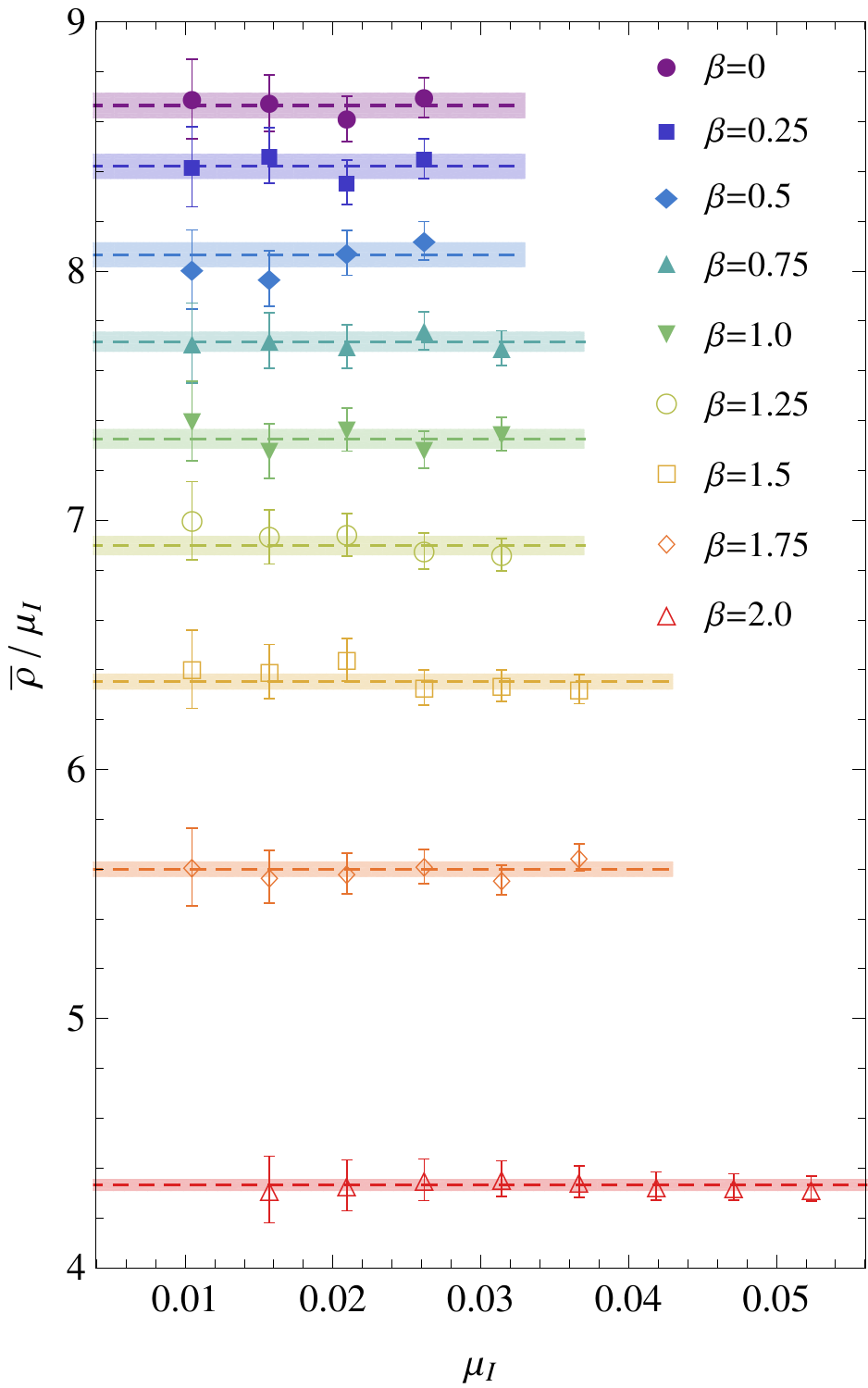}
~~~~
\includegraphics[width=6.24cm,bb=0 0 287 432]{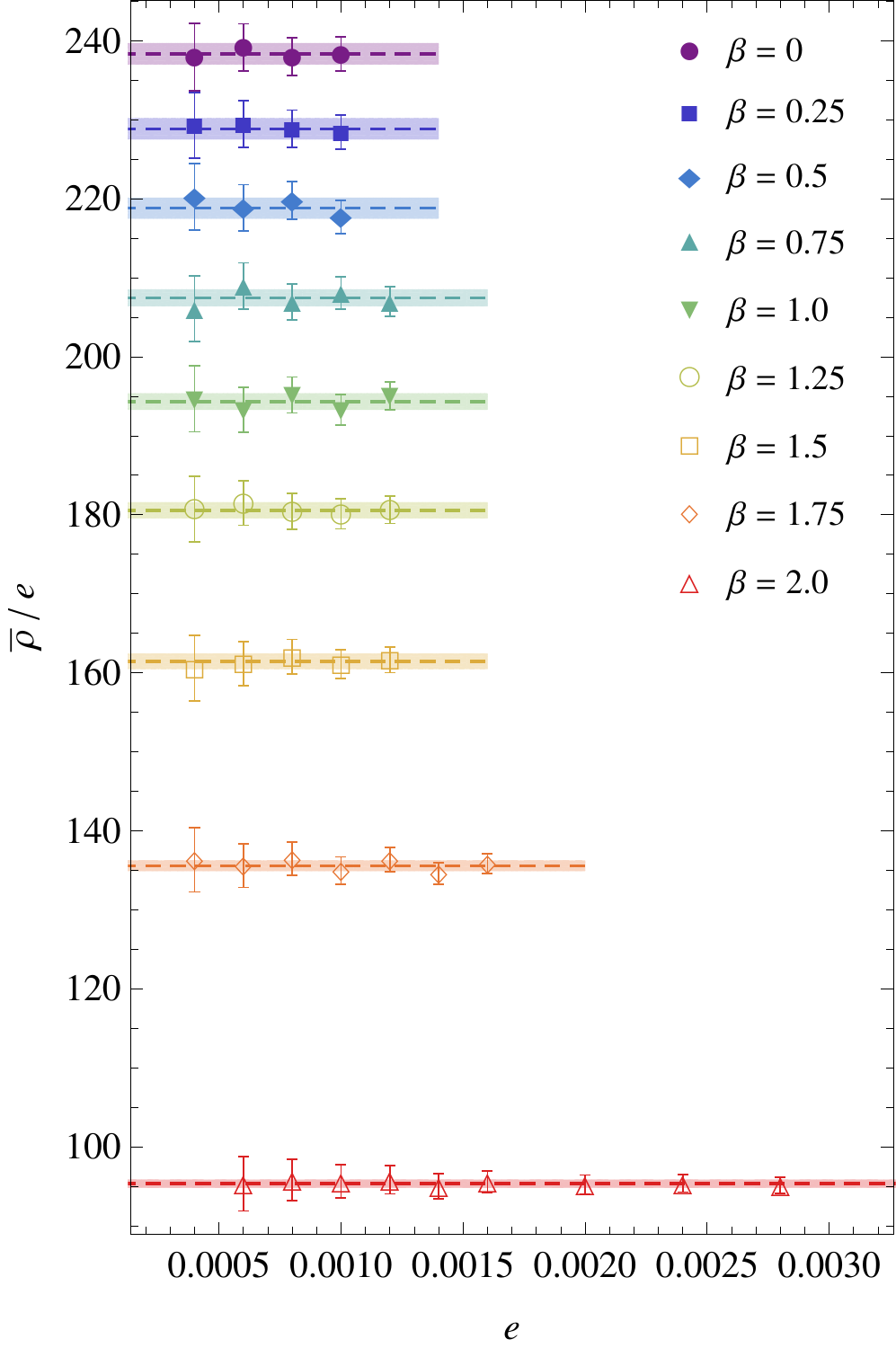}
\caption{ %
Ratios between the crossover parameter and the strength of perturbation:
$\bar{\rho} / \mu_{\mathrm{I}}$ for the SU(2)+ICP model (left) and 
$\bar{\rho} / e$ for the SU(2)$\times$U(1) model (right), on $V=6^4$ and at various $\beta$.
Colored dashed lines and strips represent the weighted averages and combined errors of the ratios. 
} %
\label{fig:SU2U1-SU2ABF_b0-20_rhoove}
\end{figure}

The pseudo-scalar decay constant $F$ 
in Eqs.~\eqref{decayconstantSU2ABF} and \eqref{decayconstantSU2U1}
is essentially the proportionality constant between the perturbation strength $\mu_{\mathrm{I}}$ or $e$ and the
crossover parameter $\rho$. 
In Fig.~\ref{fig:SU2U1-SU2ABF_b0-20_rhoove} we exhibit
the ratios $\bar{\rho}/\mu_{\mathrm{I}}$ for the SU(2)+ICP model (left)
and the $\bar{\rho}/e$ for SU(2)$\times$U(1) model (right), both on $V=6^4$ and at various $\beta$.
For each model we observe excellent stability of the ratios
as the perturbation $\mu_{\mathrm{I}}$ or $e$ is varied.
This justifies the use of combining
$\bar{\rho}/\mu_{\mathrm{I}}$ or $\bar{\rho}/e$ for all available $\mu_{\mathrm{I}}$ or $e$ at each $\beta$,
so that errors in the combined ratios are significantly reduced. 
Pseudo-scalar decay constants determined in this way 
are exhibited 
in the fourth column (SU(2)+ICP) and 
in the sixth column (SU(2)$\times$U(1))
of Table \ref{tab:low-energy-constants}.
For the SU(2)+ICP model, $\Sigma$ and $F^2$ are determined with $10^{-4}$ and $10^{-3}$ precision, respectively,
on our small lattices of $4^4$ and $6^4$.
This observation, first advocated in Ref.~\cite{DHSS05} in the context of SU(3) gauge theory with isospin ICP,
enables us to extrapolate their value to the thermodynamic limit $V\to\infty$,
listed in the bottom rows of Table \ref{tab:low-energy-constants}.
On the other hand, for the SU(2)$\times$U(1) model
the combination $F^2 \mu_{\mathrm{I}}^2/e^2$ is found to scale linearly with the lattice volume $V$.
Thus we listed the extrapolated values of $F^2 \mu_{\mathrm{I}}^2 / (e^2 V)$
in the thermodynamic limit.

\begin{figure}[H] 
\centering 
\includegraphics[width=6cm,bb=0 0 280 432]{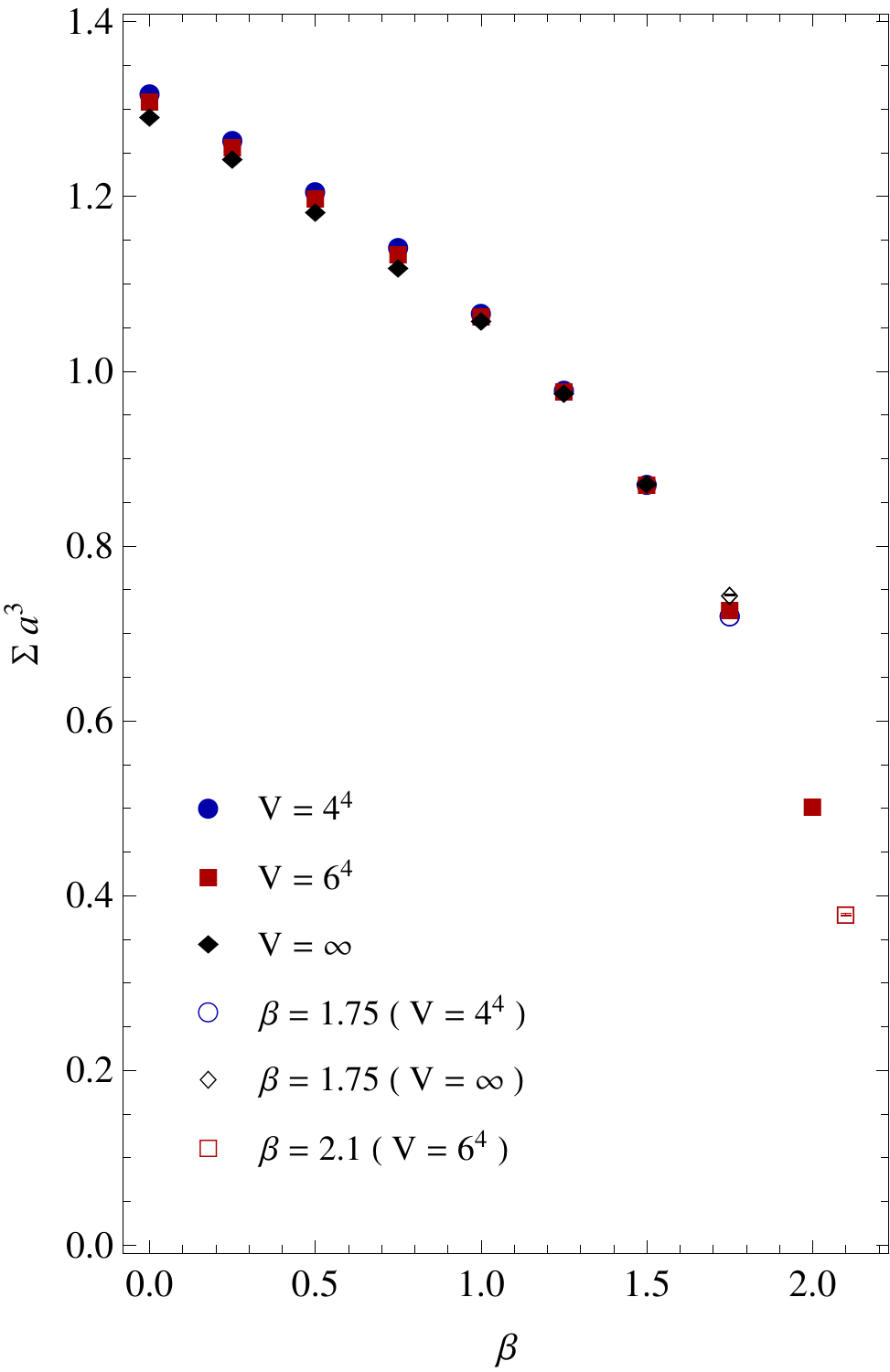}
~~
\includegraphics[width=6.2cm,bb=0 0 287 432]{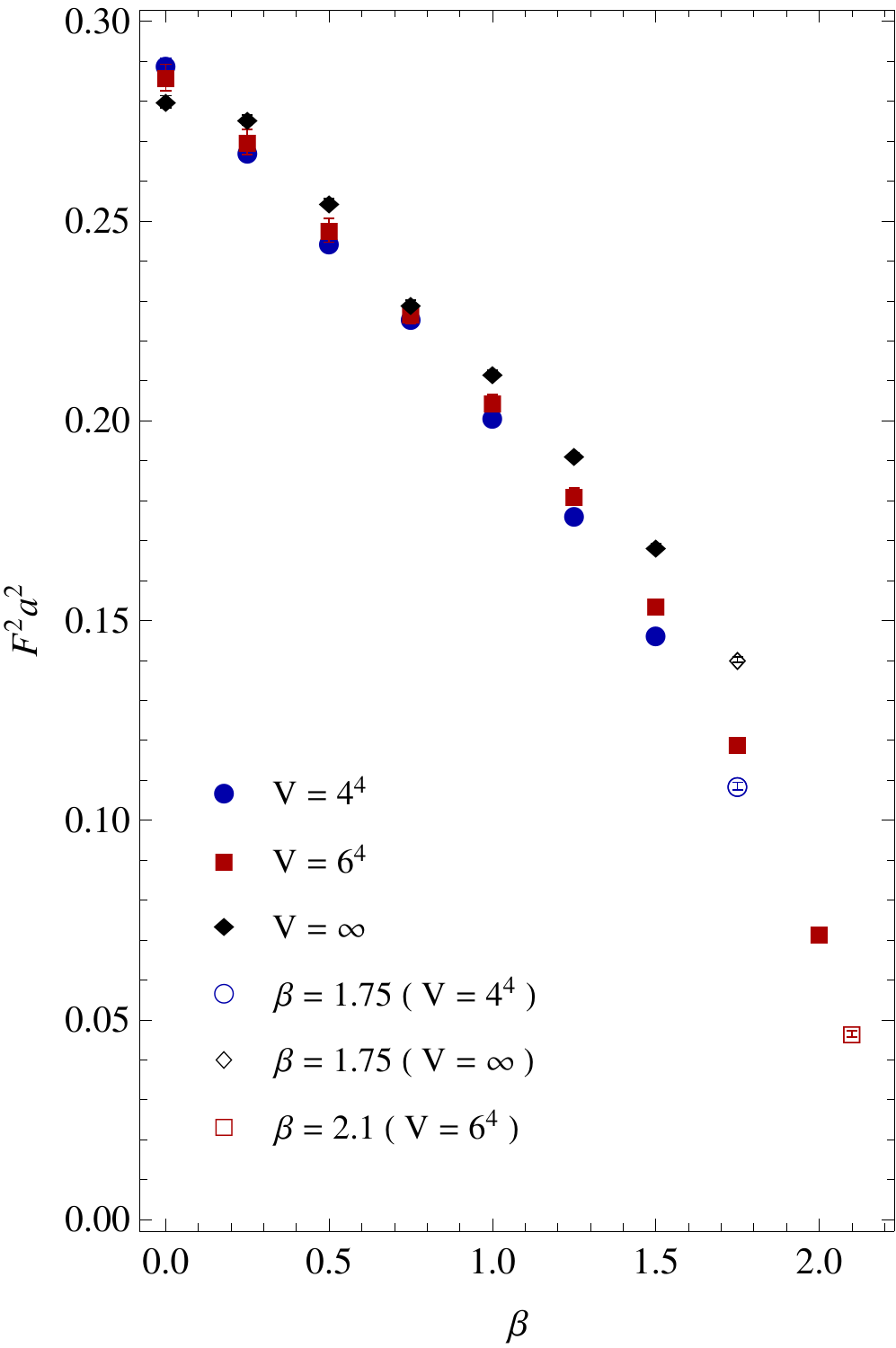}
\ \\
\ \\
\includegraphics[width=6cm,bb=0 0 280 432]{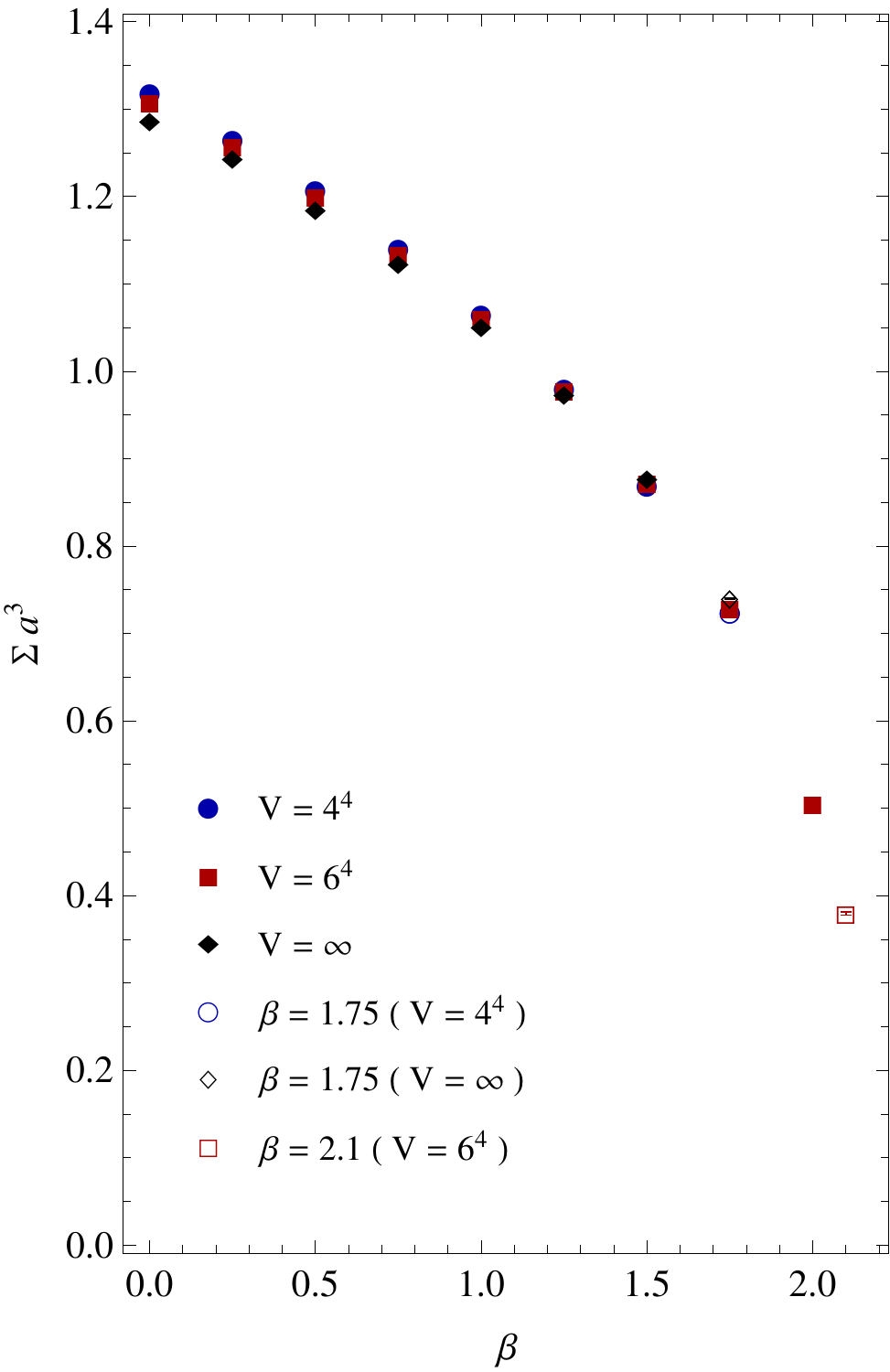}
~~
\includegraphics[width=6.2cm,bb=0 0 289 431]{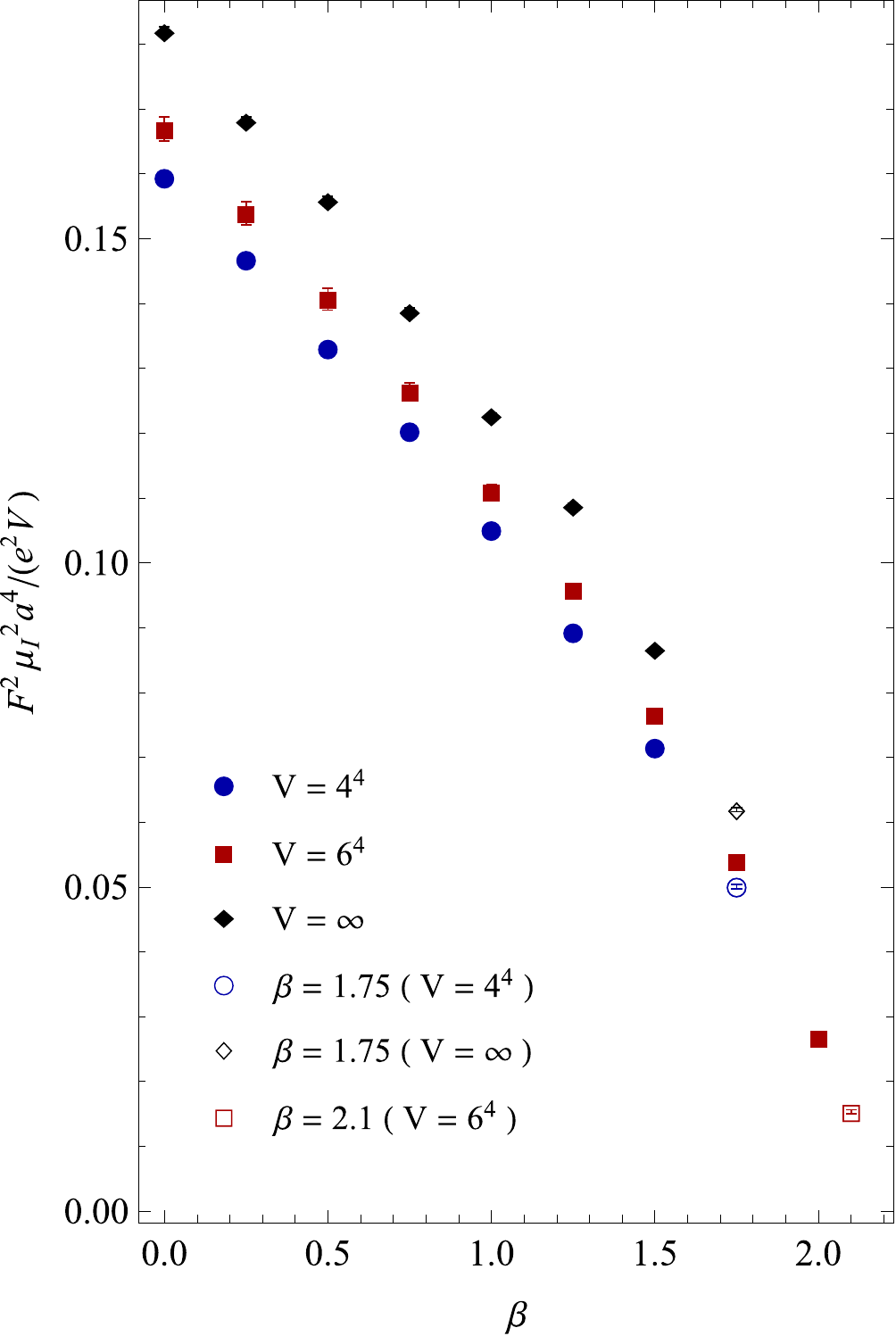}
\caption{ %
Coupling dependence of low-energy constants $\Sigma$ (left) and $F^2$ (right)
derived from the SU(2)+ICP model (top) and 
from the SU(2)$\times$U(1) model (bottom). 
Blue and red filled symbols represent the values on $V=4^4$ and $6^4$,
and black filled symbols their extrapolated values to $V\to\infty$.
Empty symbols correspond to the cases marked $*$ in Tables 1--7
(using limited ranges for fitting $P_k(s)$)
and extrapolations thereof.
}
\label{fig:low-energy-constants}
\end{figure}

Finally, in Fig.~\ref{fig:low-energy-constants}
we exhibit SU(2)-coupling dependences of the low-energy constants on $V=4^4, 6^4,$ and $\infty$
in the SU(2)+ICP model (top) and in the SU(2)$\times$U(1) model. 
The error bars are so small that they are almost obscured by the symbols.

\begin{table}[H] 
\caption{ %
Low-energy constants $\Sigma$ and $F$ 
derived from the SU(2)+ICP model 
and SU(2)$\times$U(1) model on $V=4^4$, $6^6$, and their extrapolation to the thermodynamic limit (TDL).
} %
\label{tab:low-energy-constants} 
\centering 
\begin{tabular}{ll|ll|lcl} \hline \hline 
~ & ~ & \multicolumn{2}{c|}{SU(2)+ICP} & \multicolumn{3}{c}{SU(2)$\times$U(1)}  \\ \hline  
$V$ & $\beta$ 
& \hspace{12.5pt}$\Sigma a^3$ & \hspace{10pt}$F^2 a^2$ 
& \hspace{12.5pt}$\Sigma a^3$ & $F^2 \mu_{\mathrm{I}}^2 a^4 / e^2$ & $F^2 \mu_{\mathrm{I}}^2 a^4 / (e^2 V)$ 
 \\ 
\hline 
$4^4$ & 0     & 1.318\hspace{1.5pt}0(4) & 0.289(2)  & 1.317\hspace{1.5pt}9(4) & 40.8(2) & \hspace{10pt}0.159\hspace{1.5pt}4(8)  \\  
      &    0.25   & 1.264\hspace{1.5pt}5(4) & 0.267(1)  & 1.264\hspace{1.5pt}1(4) & 37.6(2) & \hspace{10pt}0.146\hspace{1.5pt}8(7)  \\  
      &    0.5     & 1.205\hspace{1.5pt}7(4) & 0.244(1) & 1.207\hspace{1.5pt}0(4) & 34.1(2) & \hspace{10pt}0.133\hspace{1.5pt}1(7)   \\  
      &    0.75   & 1.141\hspace{1.5pt}6(4) & 0.226(1)  & 1.139\hspace{1.5pt}5(4) & 30.8(2) & \hspace{10pt}0.120\hspace{1.5pt}3(6) \\  
      &    1.0     & 1.066\hspace{1.5pt}9(5) & 0.201(1)  & 1.064\hspace{1.5pt}9(5) & 26.9(1) & \hspace{10pt}0.105\hspace{1.5pt}1(6) \\  
      &    1.25   & 0.978\hspace{1.5pt}3(7) & 0.176(1)  & 0.979\hspace{1.5pt}8(7) & 22.9(1)  & \hspace{10pt}0.089\hspace{1.5pt}3(4) \\  
      &    1.5     & 0.871\hspace{1.5pt}3(2) & 0.146\hspace{1.5pt}3(9) & 0.869(2)   & 18.3(1) & \hspace{10pt}0.071\hspace{1.5pt}5(4) \\  
      &    1.75$^*$  & 0.720(2) & 0.109(1) & 0.723(1)  & 12.8(1) & \hspace{10pt}0.050\hspace{1.5pt}0(4) \\  
\hline \hline
$6^4$ & 0     & 1.309\hspace{1.5pt}2(8) & 0.286(3) & 1.307\hspace{1.5pt}5(8) & 216(2) & \hspace{10pt}0.167(2)   \\  
      &    0.25   & 1.257\hspace{1.5pt}4(8) & 0.270(3) & 1.257\hspace{1.5pt}3(8) & 199(2) & \hspace{10pt}0.154(2)   \\  
      &    0.5     & 1.198\hspace{1.5pt}2(9) & 0.248(3)  & 1.199\hspace{1.5pt}6(8) & 182(2) & \hspace{10pt}0.141(2)  \\  
      &    0.75   & 1.134\hspace{1.5pt}2(7) & 0.227(2)  & 1.134\hspace{1.5pt}1(7) & 164(2) & \hspace{10pt}0.126(1)  \\  
      &    1.0     & 1.063\hspace{1.5pt}9(7) & 0.204(2)  & 1.060\hspace{1.5pt}4(7) & 144(2) & \hspace{10pt}0.111(1)  \\  
      &    1.25   & 0.977\hspace{1.5pt}4(6) & 0.181(2) & 0.977\hspace{1.5pt}6(6) & 124(1) & \hspace{10pt}0.096(1)   \\  
      &    1.5     & 0.871\hspace{1.5pt}4(6) & 0.154(2)  & 0.871\hspace{1.5pt}6(6) & 99(1) & \hspace{10pt}0.076\hspace{1.5pt}6(9)  \\  
      &    1.75   & 0.727\hspace{1.5pt}9(7) & 0.119(1)  & 0.728\hspace{1.5pt}6(7) & 70.0(7) & \hspace{10pt}0.054\hspace{1.5pt}0(5) \\  
      &    2.0     & 0.502(2)  & 0.071\hspace{1.5pt}5(8)  & 0.505(2)  & 34.6(4) & \hspace{10pt}0.026\hspace{1.5pt}7(3) \\ 
      &    2.1$^*$ & 0.379(1)  & 0.046\hspace{1.5pt}5(9)  & 0.379(2)  & 19.8(4) & \hspace{10pt}0.015\hspace{1.5pt}3(3) \\  
\hline \hline 
TDL &    0      & 1.291\hspace{1.5pt}6(4) & 0.280(2)  & 1.286\hspace{1.5pt}5(4)   & --         & \hspace{10pt}0.181\hspace{1.5pt}9(7)   \\  
      &    0.25   & 1.243\hspace{1.5pt}2(4) & 0.275(1) & 1.243\hspace{1.5pt}5(4)   & --         & \hspace{10pt}0.168\hspace{1.5pt}1(7)    \\  
      &    0.5     & 1.183\hspace{1.5pt}2(4) & 0.254(1) & 1.184\hspace{1.5pt}8(3)  & --         & \hspace{10pt}0.155\hspace{1.5pt}9(6)   \\  
      &    0.75   & 1.119\hspace{1.5pt}5(3) & 0.229(1) & 1.123\hspace{1.5pt}4(3)   & --         & \hspace{10pt}0.138\hspace{1.5pt}7(6)   \\  
      &    1.0     & 1.057\hspace{1.5pt}9(4) & 0.212(1)  & 1.051\hspace{1.5pt}3(4)  & --         & \hspace{10pt}0.122\hspace{1.5pt}7(4)  \\  
      &    1.25   & 0.975\hspace{1.5pt}6(5) & 0.191\hspace{1.5pt}2(9)  & 0.973\hspace{1.5pt}2(5)  & --         & \hspace{10pt}0.108\hspace{1.5pt}8(4)  \\  
      &    1.5     & 0.871\hspace{1.5pt}7(5) & 0.168\hspace{1.5pt}4(8)  & 0.876\hspace{1.5pt}9(5)  & --         & \hspace{10pt}0.086\hspace{1.5pt}7(4)  \\ 
      &    1.75$^*$  & 0.743\hspace{1.5pt}9(6) & 0.140\hspace{1.5pt}2(7)  & 0.739\hspace{1.5pt}9(6)  & --         & \hspace{10pt}0.061\hspace{1.5pt}9(3)  \\  
\hline \hline 
\end{tabular}
\end{table}%

\section{Conclusions}
We have analytically evaluated the $k$th smallest eigenvalue distributions $p_k(s)$
for a random matrix ensemble interpolating chGSE and chGUE 
using a Nystr\"{o}m-type method applied to the Fredholm Pfaffian and resolvents of 
the quaternion kernel.
These random matrix results are applied to fit
the spectra of fundamental, staggered Dirac operators of 
SU(2) gauge theory with imaginary chemical potential and
of SU(2)$\times$U(1) gauge theory on small lattices,
from the strong-coupling to the near-scaling regions.
Combined use of the first four nondegenerate Dirac eigenvalue distributions 
in place of the spectral density or the smallest eigenvalue distribution
of unperturbed SU(2) gauge theory 
enables us to determine the chiral condensate with $\mathcal{O}(10^{-4})$ precision.
Excellent one-parameter fitting of $\chi^2/\text{d.o.f.}<2$
between {\em non-cumulative} individual distributions and eigenvalue histograms 
is achieved for almost all cases of U(1) perturbations.
Combined use of the first four eigenvalue distributions 
also contributed to a reduction of the errors in the crossover parameter $\rho$.
The acute sensitivity of $p_k(s)$ on $\rho$, and
the observed linear dependence of 
$\rho$ on the perturbation strength (AB flux $\varphi$ or U(1) coupling $e$) resulted in
determination of the pseudo-scalar
decay constant $F$ (i.e., the coefficient of the pseudoreality-breaking term) 
with $\mathcal{O}(10^{-3})$ precision.

Our method of determining $F$ in QCD-like theories, which has proved to be feasible on relatively small-sized lattices, 
is clearly advantageous over the conventional method of using axial current correlators,
which inevitably requires a large temporal dimension.
A possible application of our method
would be towards technicolor candidate gauge theories with fermions in (pseudo)real representations,
such as SU($N$) gauge theory with two adjoint flavors \cite{Hietanen09}, 
which corresponds to the chGSE class if simulated with an overlap Dirac operator \cite{Ver94}.
Hyper-precise determination of its ``low-energy'' constants (i.e.,~Higgs couplings) from lattice simulations 
using our ICP method would,
upon comparison with knowledge from collider experiments,
contribute to single out a credible scenario from such BSM candidates.

\section*{Acknowledgements}
S.M.N. thanks J.~Verbaarschot and P.~Forrester for valuable discussions.
This work is supported in part by JSPS Grants-in-Aids for Scientific Research (C) Nos. 25400259 and 17K05416.

\end{document}